 \documentclass{aa} 

\usepackage{graphicx}
\usepackage{txfonts}
\usepackage[usenames]{color} 
\usepackage{tabularx} 
\renewcommand{\arraystretch}{1.5} 
\usepackage{footnote}
\makesavenoteenv{table*}
\usepackage{rotating}
\usepackage{float}
\usepackage{longtable} 
\usepackage{lscape} 

\usepackage{array}
\newcolumntype{H}{>{\setbox0=\hbox\bgroup}c<{\egroup}@{}}
\usepackage{multirow}
\usepackage{scalerel}
\usepackage{tikz}
\usetikzlibrary{svg.path}
\definecolor{orcidlogocol}{HTML}{A6CE39}
\tikzset{
  orcidlogo/.pic={
    \fill[orcidlogocol] svg{M256,128c0,70.7-57.3,128-128,128C57.3,256,0,198.7,0,128C0,57.3,57.3,0,128,0C198.7,0,256,57.3,256,128z};
    \fill[white] svg{M86.3,186.2H70.9V79.1h15.4v48.4V186.2z}
                 svg{M108.9,79.1h41.6c39.6,0,57,28.3,57,53.6c0,27.5-21.5,53.6-56.8,53.6h-41.8V79.1z M124.3,172.4h24.5c34.9,0,42.9-26.5,42.9-39.7c0-21.5-13.7-39.7-43.7-39.7h-23.7V172.4z}
                 svg{M88.7,56.8c0,5.5-4.5,10.1-10.1,10.1c-5.6,0-10.1-4.6-10.1-10.1c0-5.6,4.5-10.1,10.1-10.1C84.2,46.7,88.7,51.3,88.7,56.8z};
  }
}

\newcommand\orcid[1]{\href{https://orcid.org/#1}{\mbox{\scalerel*{
\begin{tikzpicture}[yscale=-1,transform shape]
\pic{orcidlogo};
\end{tikzpicture}
}{|}}} \href{#1}{#1}}
\usepackage[breaklinks=true,colorlinks, linkcolor=blue,urlcolor=blue,citecolor=blue]{hyperref}

\usepackage{threeparttablex} 
 
\usepackage[switch]{lineno}
\usepackage[normalem]{ulem}
\usepackage{xcolor}

\begin{document}
\newcommand\redsout{\bgroup\markoverwith{\textcolor{red}{\rule[0.5ex]{2pt}{1.0pt}}}\ULon}

   \title{Solar activity relations in energetic electron events measured by the MESSENGER mission}

   \subtitle{}

   \author{L. Rodríguez-García
          \inst{1}
          \and L. A. Balmaceda\inst{2,3}
          \and
          R. Gómez-Herrero\inst{1}
          \and A. Kouloumvakos\inst{4}\and  N. Dresing\inst{5}
          \and D. Lario\inst{3}
          \and\newline I. Zouganelis\inst{6}
          \and  A. Fedeli\inst{5}
            \and F. Espinosa Lara\inst{1} 
          \and I. Cernuda\inst{1}
          \and G. C. Ho\inst{4}
          \and R. F. Wimmer-Schweingruber\inst{7}
          \and \newline  J. Rodríguez-Pacheco\inst{1}
          }
   \institute{Universidad de Alcalá, Space Research Group (SRG-UAH), Plaza de San Diego s/n, 28801 Alcalá de Henares, Madrid, Spain \\
              \email{l.rodriguezgarcia@edu.uah.es}
         \and Heliophysics Science Division, NASA Goddard Space Flight Center, Greenbelt, MD 20771, USA 
             \and Physics and Astronomy Department, George Mason University, 4400 University Drive, Fairfax, VA 22030, USA \and The Johns Hopkins University Applied Physics Laboratory, 11101 Johns Hopkins Road, Laurel, MD 20723, USA
             \and Department of Physics and Astronomy, University of Turku, 20014 Turku, Finland\and European Space Agency (ESA), European Space Astronomy Centre (ESAC), Camino Bajo del Castillo s/n, 28692 Villanueva de
la Cañada, Madrid, Spain\and
             Institut fuer Experimentelle und Angewandte Physik, University of Kiel, Leibnizstrasse 11, 24118 Kiel, Germany
            }

   \date{Received December 1, 2022; Accepted March 21, 2023}

 
  \abstract
  {}
   {We perform a statistical study of the relations between the properties of solar energetic electron (SEE) events measured by the MESSENGER mission from 2010 to 2015 and the parameters of the respective parent solar activity phenomena to identify the potential correlations between them. During the time of analysis MESSENGER heliocentric distance varied between 0.31 and 0.47 au.   } 
   {We used the published list 
 by \cite{Rodriguez-Garcia2023_messenger} of 61 SEE events measured by MESSENGER, which  includes the information of the near-relativistic electron peak intensities, the peak-intensity energy spectral indices, and the measured X-ray peak intensity of the flares related to the SEE events. Taking advantage of multi-viewpoint remote sensing observations, we reconstructed, whenever possible, the associated coronal mass ejections (CMEs) and shock waves; and we determined the 3D properties (location, speed, and width) of the CMEs and the maximum speed of the 3D CME-driven shocks in the corona. We used different methods  (Spearman, Pearson, and a Bayesian approach, namely the Kelly method to linear regression) to estimate the correlation coefficients between the flare intensity, maximum speed at the apex of the CME-driven shock, CME speed at the apex, and CME width with the electron peak intensities and with the energy spectral indices. In this statistical study, we considered and addressed the limitations of the particle instrument on board MESSENGER (elevated background intensity level, anti-Sun pointing). } 
   {There is an asymmetry to the east in the range of connection angles (CAs) for which the SEE events present the highest peak intensities, where the CA is the longitudinal separation between the footpoint of the magnetic field connecting to the spacecraft and the flare location. Based on this asymmetry, we define the subsample of well-connected events as when -65$^{\circ}\leq$ CA $\leq+33^{\circ}$. For the well-connected sample, we find moderate to strong correlations between the near-relativistic electron peak intensity and the 3D CME-driven shock maximum speed at the apex (Spearman: cc=0.53$\pm$0.05; Pearson: cc=0.65$\pm$0.04; Kelly: cc=0.87$\pm$0.20), the flare peak intensity (Spearman: cc=0.63$\pm$0.03; Pearson: cc=0.59$\pm$0.03; Kelly: cc=0.74$\pm$0.30), and the 3D CME speed at the apex (Spearman: cc=0.50$\pm$0.04; Pearson: cc=0.46$\pm$0.03; Kelly: cc=0.60$\pm$0.39). When including poorly-connected events (full sample), the relations between the peak intensities and the solar activity phenomena are blurred, showing lower correlation coefficients.   }
   {Based on the comparison of the correlation coefficients presented in this study using near 0.4 au data, (1) both flare and shock-related processes may contribute to the acceleration of near relativistic electrons in large SEE events, in agreement with previous studies based on near 1 au data; and (2) the maximum speed of the CME-driven shock is a better parameter to investigate particle acceleration related mechanisms than the average CME speed, as suggested by the stronger correlation with the SEE peak intensities.
}
   
   \titlerunning{Solar activity relations in energetic electron events measured by the MESSENGER mission} \authorrunning{L. Rodríguez-García et al.}
   \maketitle
%
\section{Introduction}
\label{sec:Introduc}

Solar energetic electron (SEE) events are sporadic enhancements of electron intensities associated to solar transient activity. In the inner heliosphere, these intensity enhancements are usually measured in situ at near-relativistic ($\gtrsim$ 30 keV) and relativistic ($\gtrsim$ 0.3 MeV) energies. The mechanisms proposed to explain the origin of solar near-relativistic electron events include: (1) acceleration during magnetic reconnection processes associated to solar jets \citep{Krucker2011} and flares \citep{2007Kahler}; (2) acceleration during magnetic restructuring in the aftermath of coronal mass ejections (CMEs) and in the current sheets formed at the wake of CMEs \citep[e.g.][]{Kahler1992, 2004MaiaPick, 2005Klein}; (3) and/or acceleration at shocks driven by fast CMEs \citep{2002Simnett}. 

Previous statistical studies point out that multiple acceleration processes may contribute to the acceleration of quasi-relativistic energetic electrons \citep[e.g.][]{Kouloumvakos2015,2015Trottet}. In particular, \cite{2015Trottet} concluded that near-relativistic electrons ($\sim$175 keV) in large SEP events have a mixed flare-CME origin, supported by \cite{2022Dresing} conclusions: electrons in the MeV range are mainly accelerated by CME-driven shocks, while lower energy ($\sim$50 keV) electrons are likely produced by a mixture of flare and shock-related acceleration processes.

Many efforts have been made to identify a unique accelerator by investigating the correlations between solar energetic particle (SEP) parameters, especially their peak intensity, and the properties of the associated solar activity phenomena, such as the solar flare X-ray peak intensity, CME speed and width, and CME-driven shock speed \citep[e.g.][]{2001Kahler,Richardson2014,Papaioannou2016,Kouloumvakos2019,Xie2019,Kihara2020}. The aforementioned studies are mainly based on measurements near 1 au, however particle propagation in the interplanetary space affect SEE properties. Thus, the observation of SEE events by spacecraft located at heliocentric distances less than 1 au (i.e., closer to the acceleration site) is essential to infer the mechanisms associated to their acceleration \citep[e.g.][]{2016AguedaLario}. To minimize projection effects in the CME properties and in the CME-driven shock speed, forward modelling is generally used to reconstruct the three-dimensional (3D) morphology of the CME and CME-driven shock in the corona, using imaging observations from multiple vantage points \citep[e.g.][]{Kwon2014,Kouloumvakos2016}.

In this paper we study the relationship between solar activity (flare, CME, CME-shock) and the properties of SEE events measured by the \textit{MErcury Surface Space ENvironment GEochemistry and Ranging} \citep[MESSENGER;][]{Solomon2007MESSENGER} mission near 0.3 au, presented by \cite{Rodriguez-Garcia2023_messenger}, hereafter Paper I.  In particular, we use energetic electron measurements from 2010 February to 2015 April when
MESSENGER heliocentric distance varied between 0.31 and 0.47 au. We take advantage of the good remote-sensing coverage from near 1 au spacecraft, such as the twin spacecraft of the \textit{Solar TErrestrial RElations Observatory} \citep[STEREO;][]{Kaiser2008STEREO} and the \textit{SOlar and Heliospheric Observatory} \citep[SOHO;][]{Domingo1995SOHO}, to reconstruct the 3D CMEs and CME-driven shocks associated to the SEE events. These multi-point observations allow us to study the relations between the solar source parameters and the peak intensity and peak-intensity energy spectrum of SEE events closer to the Sun.  

Thus, the main goal of the study is to relate the SEE peak intensities and peak-intensity energy spectra to various parameters of the parent solar activity, presented in Sect. \ref{sec:parent_activity_relations}. The rest of the paper is structured as follows.  The instrumentation used in this study is introduced in Sect. \ref{sec:Instrumentation}.  A summary of the SEE events measured by MESSENGER that were presented in Paper I is shown in Sect. \ref{sec:SEP_measured_by_MESSENGER}. We include in Sect. \ref{sec:solar_parent_activity} the 3D reconstructions of the CMEs and CME-driven shocks related to the SEE events. Section \ref{sec:summary_discussion} summarizes and discusses the main findings of the study. 
\section{Instrumentation} 
\label{sec:Instrumentation}

The statistical study of the relations between SEE events and their parent solar source requires the analysis of both remote-sensing and in situ data from a wide range of instrumentation on board different spacecraft. We used data from MESSENGER, STEREO, SOHO, the \textit{Solar Dynamics Observatory} \citep[SDO;][]{Pesnell2012}, and the \textit{Geostationary Operational Environmental Satellites} \citep[GOES;][]{Garcia1994}. 

Remote-sensing observations of CMEs and related solar activity phenomena on the Sun's surface were provided by the Atmospheric Imaging Assembly \citep[AIA;][]{Lemen2012} on board SDO, the C2 and C3 coronagraphs of the Large Angle and Spectrometric COronagraph \citep[LASCO;][]{Brueckner1995} instrument on board SOHO, and the Sun Earth Connection Coronal and Heliospheric Investigation \citep[SECCHI;][]{Howard2008SECCHI} instrument suite on board STEREO. In particular, we used the COR1 and COR2 coronagraphs and the Extreme Ultraviolet Imager \citep[EUVI;][]{Wuelser2004}, part of the SECCHI suite. 

In Paper I, data from the X-Ray telescopes of the GOES satellites\footnote{\url{https://satdat.ngdc.noaa.gov/sem/goes/data/avg/}\label{goes_data}} and in situ energetic particle observations provided by the Energetic Particle Spectrometer (EPS), part of the Energetic Particle and Plasma Spectrometer \citep[EPPS;][]{Andrews2007EPPS} on board MESSENGER, were used.

\section{SEE events measured by MESSENGER}
\label{sec:SEP_measured_by_MESSENGER}
The SEE events included in this study are presented in Paper I, where the data source and selection criteria are explained in detail. We summarize here the more relevant information.

\subsection{Data source and SEE event selection criteria}
\label{sec:Data_sources_SEP_selection_criteria}

The study includes MESSENGER data from 2010 February 7 to 2015 April 30. In this period, coinciding with most of the rising, maximum, and early decay phase of solar cycle 24, MESSENGER heliocentric distance varied from 0.31 to 0.47 au. 
 
The EPS instrument on board MESSENGER measured electrons from $\sim$25 keV to $\sim$1 MeV. The electron energies chosen in Paper I for the SEE event identification and statistical analysis were 71-112 keV. In the case of the analysis of energy spectra, the energies used were from $\sim$71 keV to $\sim$1 MeV divided into six energy bins. The EPS instrument was mounted on the far-side of the spacecraft, with a field of view divided into six sectors pointing in the antisunward direction, so it mostly detected particles moving sunward. Usually, SEP events present a higher particle flux and earlier onset in the sunward-pointing telescope that is aligned with the IP magnetic field \citep[e.g.][]{1991Kunow}. Therefore, MESSENGER observations presumably provide a lower limit to the actual peak intensities of the SEE events and an upper limit to the timing of occurrence of such peaks.

The peak intensity in the prompt component of the event, namely the maximum intensity reached shortly (usually $\lesssim$6 hours) after its onset, was chosen as the maximum intensity. Although electron intensity enhancements associated to the passage of IP shocks are rare \citep{2003Lario,2016Dresing}, by selecting the prompt component of the SEE events, the possible effect that traveling IP shocks might have on the continuous injection of particles was minimized. Therefore, the peak intensity of the SEE events was observed when the associated CMEs were still close to the Sun. 

Due to the elevated background level of the EPS instrument, the selected events showed intensities that are normally above $\sim$10\textsuperscript{4} (cm\textsuperscript{2} sr s MeV)\textsuperscript{-1}. An exception to this is the period of 2011 August, when EPS geometric factor was modified allowing for a temporary detection of less intense events. In order to keep the self-consistency of the analysis, events number 6 and 7, measured in August 2011 during the period of increased geometric factor of the MESSENGER/EPS instrument, were not included in this study. 

\subsection{MESSENGER SEE event list}
\label{sec:MESSENGER SEE list}

Table \ref{Table:SEP_list} shows the list of the 61 SEE events presented in Paper I.  Columns 1-3 identify each SEE event with a number (1), the solar event date (2), and the time of the type III radio burst onset (3). The symbol (ˆ) is used to indicate when the type III burst onset time is uncertain due to occultation or multiple radio emissions at the same time during the onset of the event. Column 4 provides the location of the solar flare in Stonyhurst coordinates, either identified in Paper I or consulted in different catalogues and studies (table 2 of Paper I). The flare class indicated in square brackets is based on the 1-8 {\AA} channel measurements of the X-Ray telescopes on board GOES. 
To be consistent with previous statistical studies \citep[e.g.][]{Richardson2014}, the flare location was chosen as the site of the putative particle source. Columns 5-7 are described in Sect. \ref{sec:solar_parent_activity}.

Column 8 in Table \ref{Table:SEP_list} shows the MESSENGER connection angle (CA), which is the longitudinal separation between the flare site location and the footpoint of the magnetic field line connecting to the spacecraft, based on a nominal Parker spiral, as discussed below. Positive CA denotes a flare source located at the western side of the spacecraft's magnetic footpoint. The magnetic footpoint for MESSENGER was estimated assuming a Parker spiral with a constant speed of 400 km s\textsuperscript{-1} using the \textit{Solar-MACH tool}  available online\footnote{\url{https://doi.org/10.5281/zenodo.7100482}\label{Solar-Mach}} \citep{Gieseler2022}, as MESSENGER lacks solar wind measurements. The heliocentric distance of the MESSENGER spacecraft at the time of the event is given in Col. 9, which varied between 0.31 au and 0.47 au during the time interval considered in the study. Column 10 summarizes the 71-112 keV electron peak intensities corresponding to the prompt component of the event as discussed above. The pre-event background level is given in parentheses. 
 
An event was considered widespread when either the MESSENGER |CA| is more than 80$^{\circ}$ or the longitudinal separation between MESSENGER and another spacecraft near 1 au that detected the event was more than 80$^{\circ}$ \citep{Dresing2014}. We indicate these events with an asterisk next to the event number in Col. 1 of Table \ref{Table:SEP_list}. A total of 44 SEE events can be characterized as widespread according to our criteria. However, the number of widespread events could be larger since, apart from not sampling all the heliolongitudes with the existing constellation of spacecraft, there were events with a high prior-event-related background or with no data available for some of the spacecraft, so no particle increase could be measured. 
 
As detailed in Sect.~\ref{sec:solar_parent_activity}, we found a CME (CME-driven shock) related to the electron increase in 57 (56) events. For these associations we previewed the available conoragraphic data from SOHO/LASCO or STEREO/COR2 near the flare and SEE onset times and register the related events. In almost all the cases, the CMEs and CME-driven shock waves were very prominent and clearly related to the flare eruption. Relativistic ($\sim$1~MeV) electron intensity enhancements were observed in 37 events, as indicated with a dagger in Column 11 of the list. Thus, the majority of the events detected by MESSENGER are CME- and CME-driven shock-related events, with a high peak intensity level and the presence of $\sim$1~MeV electrons, which were observed by widely separated spacecraft. The observed characteristics of the SEE events are expected due to the high background level of MESSENGER/EPS that prevents the instrument from measuring less intense events \citep[e.g. figure 1 in][]{Lario2013}.

 \section{Solar parent activity in SEE events measured by MESSENGER}
 \label{sec:solar_parent_activity}

To investigate the relations between the properties of the SEE events measured by MESSENGER and some of the parent solar source parameters, we used the flare peak intensity  measurements presented in Paper I for the events originating on the visible side of the Sun from Earth's point of view. The flare class based on GOES Soft X-ray (SXR) peak flux is given in square brackets in Col. 4 of Table \ref{Table:SEP_list}. For the far-side event number 36 (2013/08/19), the equivalent GOES intensity of the flare is given using the STEREO/EUVI light curve \citep{Nitta2013}, as explained in \cite{2021Rodriguez-Garcia} and indicated with (§) in Col. 4. The uncertainty of the logarithm of the flare intensity is estimated to be 0.1, taken as the rounding error of the measurements. 
 
In this study we performed the 3D reconstruction of the associated CMEs and CME-driven shocks for 57 and 54 SEE events, respectively.  In two events where a CME-driven shock was observed, we did not perform the 3D reconstruction as we could not trace the shock accurately. By determining the CME parameters, such as the width and speed, and the CME-driven shock speed from the 3D reconstruction, we reduced the projection effects and the final values are more accurate. Previous studies \citep[e.g.][]{Kouloumvakos2019,Xie2019,2022Dresing} show that when using in the statistical analysis the reconstructed parameters, instead of the plane-of-sky values, the estimated correlations are stronger. The reconstruction process is explained below.

\subsection{3D CME parameters}
\label{3D CME parameters}
 
 We took advantage of the multi-view spacecraft observations and reconstructed the 3D CME using the graduated cylindrical shell (GCS) model \citep{Thernisien2006GCS, Thernisien2011}. The GCS model uses the geometry of what looks like a hollow `croissant' to fit a flux-rope structure using coronagraph images from multiple viewpoints. The deviations in the parameters of the GCS analysis are given in table 2 of \cite{Thernisien2009}. The tools used for the reconstruction are (1) the \textit{rtsccguicloud.pro} routine, available as part of the \textit{scraytrace} package in the SolarSoft IDL library\footnote{\url{http://www.lmsal.com/solarsoft/}} and (2) \textit{PyThea}, a software package to reconstruct the 3D structure of CMEs and shock waves \citep{Kouloumvakos2022} written in Python and available online\footnote{\url{https://doi.org/10.5281/zenodo.5713659}\label{Pythea}}. The images underwent a basic process of calibration, and we used base-difference images to highlight the CME contour from other coronal features. As inferred from the on-disk observations of the post-eruptive loops and/or of the filament prior to the eruption, several events (10 out of 57) showed non-radial propagation or presented `curved axes'.  This last term was introduced by \cite{Rodriguez-Garcia2022CME} to refer to flux ropes that may deviate from the nominal semi-circular (croissant-like) shape and have instead an undulating axis. In these cases, the GCS parameters are chosen to better describe the portion of the CME closer to the ecliptic plane, which is closer to MESSENGER orbit plane. Then, we obtained the following 3D CME parameters from the GCS reconstruction, as detailed by \cite{Thernisien2006GCS, Thernisien2011}: (1) the \textit{half-angle}; (2) the \textit{ratio}, which sets the rate of lateral expansion of the minor radius to the height of the center of the CME at the apex; and (3) the \textit{tilt}, which is the angle of the main axis of the CME relative to the solar equator.
 
 The 3D CME speed at the apex and the CME width are given in Cols. 5 and 6 of Table \ref{Table:SEP_list}. The CME speed at the apex is derived from a linear fit of the different heights of the CME apex observed at different times taking at least three instants for the fitting. The uncertainty of the CME speed is considered to be 7\% of the value, based on \cite{Kwon2014}. The width of the CME was estimated based on \cite{Dumbovic2019}, where the semi-angular extent in the equatorial plane is represented by ${R\textsubscript{maj}-{(R\textsubscript{maj}-R\textsubscript{min})} \times |tilt|/90}$. The value of $R\textsubscript{maj}$ (face-on CME half-width) was calculated by adding $R\textsubscript{min}$ (edge-on CME half-width) to the half-angle. The $R\textsubscript{min}$ was determined as the $\arcsin(ratio)$, which is given by GCS, as presented above. The uncertainty of the CME width in the equatorial plane is taken as the deviations of the half-angle given by \cite{Thernisien2009}.

All the reconstructions were performed using three points of view (STEREO-A, -B and SDO and/or SOHO), whenever possible. For the reconstructions of event number 54 and from number 58 to 61 we only used data from the Earth point-of-view, indicated with an exclamation mark in Cols. 5 and 6 of Table \ref{Table:SEP_list}. These events occurred near the time of the solar superior conjunction of the STEREO spacecraft (from January to August 2015) and no STEREO data were available. However, these events are still included in the statistical study, keeping in mind that the reconstructed parameters could have larger uncertainties. After an exhaustive inspection of the data, we found no CME associated with events number 35, 42, 46, and 47. 

 \begin{table*}[htb]
\centering
\caption{Summary of the statistical properties of the different samples analysed in this study.}
\label{Table:statistics_samples}
\small
\tabcolsep=0.1cm
\begin{tabularx}{1\textwidth}{ccccccccccccccc} 
\hline
\hline
\multicolumn{2}{c}{Variable}&Count& Mean& STD&Min&25\%&50\%&75\%&Max&Skew&Kurt&Zskew&Zkurt&Normal\\
\hline
\multicolumn{2}{c}{(1)}&(2)&(3)&(4)&(5)&(6)&(7)&(8)&(9)&(10)&(11)&(12)&(13)&(14)\\\hline
 \multirow{2}{*}{Log Peak Int} &all&59&5.09&0.93&3.77&4.41&4.79&5.44&7.69&1.29&3.93&3.66, 0.00&1.63, 0.10&No\\
  &(w-con)&(30)&(5.41)&(1.05)&(3.77)&(4.58)&(5.21)&(6.04)&(7.69)&(0.76)&(2.71)&(1.87, 0.06)&(0.08, 0.93)&(Yes)\\\hline
 \multirow{2}{*}{Log Flare Int}&all &38&-4.5&0.75&-6.32&-5.03&-4.35&-3.93&-3.27&-0.53&2.58&-1.47, 0.14 &-0.27, 0.79&Yes \\
  &(w-con)&(18)&(-4.62)&(0.81)&(-6.32)&(-5.15)&(-4.59)&(-3.97)&(-3.27)&(-0.26)&(2.29)&(-0.57, 0.57)&(-0.41, 0.68)&(Yes)\\\hline
\multirow{2}{*}{Log Shock Spe} &all&52&3.22&0.15&2.90&3.10&3.21&3.35&3.53&0.18&2.32&0.59, 0.55&-1.14, 0.25&Yes\\
 &(w-con\textsuperscript{1})&(24)&(3.23)&(0.15)&(3.04)&(3.13)&(3.17)&(3.27)&(3.53)&(0.92)&(2.69)&(2.04, 0.04)&(0.12, 0.90)&(Yes)
\\\hline
\multirow{2}{*}{Log CME Spe} &all&55&3.08&0.20&2.41&2.96&3.11&3.23&3.43&-0.81&3.9&-2.47, 0.01&1.57, 0.11&Yes\\
 &(w-con)&(26)&(3.07)&(0.22)&(2.41)&(2.97)&(3.08)&(3.19)&(3.43)&(-0.97)&(4.62)&(-2.19, 0.03)&(1.98, 0.05)&(Yes)\\\hline

\multirow{2}{*}{Log CME Wid} &all&55&1.83&0.15&1.36&1.74&1.84&1.89&2.20&-0.25&4.17&-0.83, 0.40& 1.83, 0.07&Yes\\
 &(w-con)&(26)&(1.85)&(0.14)&(1.66)&(1.75)&(1.82)&(1.89)&(2.20)&(0.95)&(3.20)&(2.15, 0.03)&(0.79, 0.43)&(Yes)\\\hline

\multirow{2}{*}{$\delta\textsubscript{200}$} &all&42&-1.94&0.19&-2.56&-1.99&-1.92&-1.82&-1.55&-1.11&5.26&-2.88, 0.00&2.47, 0.01&No\\
 &(w-con)&(23)&(-1.97)&(0.14)&(-2.41)&(-2.01)&(-1.94)&(-1.90)&(-1.81)&(-1.85)&(6.64)&(-3.47, 0.00)&(2.88, 0.00)&(No)
\\

 \hline
\end{tabularx}
\begin{flushleft}

 \footnotesize{ \textbf{Notes.} Column 1: Variables for the full sample and well-connected events (in parenthesis, namely -65$^{\circ}$ $\leq$ connection angle $\leq$ 33$^{\circ}$) in the respectively following units: peak intensity (cm\textsuperscript{2} sr s MeV)\textsuperscript{-1}, flare intensity (W m\textsuperscript{-2}), CME-driven shock speed (km s\textsuperscript{-1}), CME speed (km s\textsuperscript{-1}), CME width (deg), spectral index (-). Column 2: total number of entries. Column 3: Average of all entries. Column 4: Standard deviation. Column 5: minimum value. Columns 6-8: 25, 50 (median), and 75 percentile mark, respectively. Column 9: maximum value. Column 10: skewness, namely the measure of the lack of symmetry. Column 11: kurtosis, namely the measure of whether the data are heavy-tailed or light-tailed relative to a normal distribution. Column 12: Z-skewness: statistics, p-value. Column 
13: Z-kurtosis: statistics, p-value. Column 14: Whether data can be considered normally distributed or not based on several criteria: visual inspection of the distribution, Z-value \cite[both skewness and kurtosis;][]{west1995structural}, Normality test \citep[not shown;][]{D'AgostinoPearson1976normalitytest}, and Anderson-Darling test \citep[not shown;][]{Sthephens1974andersontest}. \textsuperscript{1} in Col. 1: Only the Anderson-Darling test \citep[not shown;][]{Sthephens1974andersontest} is not fulfilled.}

\end{flushleft}
\end{table*}

\subsection{3D CME-driven shock speed}
\label{3D shock speed}

We also performed a reconstruction of the coronal shock waves associated to the SEE events, fitting an ellipsoid shape to the observations, although the actual shape of the outermost wave usually observed in front of the CME probably may differ from the assumed ideal contour. In order to do this, we used the \textit{PyThea}\textsuperscript{\ref{Pythea}} tool, which applies the ellipsoid model developed by \cite{Kwon2014} to quasi-simultaneous images from different vantage points. In this case, we used running-difference images to highlight the shock front in the calibrated images. The fitting process is explained in detail by \cite{Kwon2014} and by \cite{Kouloumvakos2019}. We found no CME-driven shock for events number 30, 35, 42, 46 and 47. They are related to one slow CME (event number 30) and the four CME-less events discussed above.  In events number 58 to 60 we used only data from the Earth point-of-view, indicated with an exclamation mark in Col. 7 of Table \ref{Table:SEP_list}, due to the lack of STEREO imaging during the solar superior conjunction, as discussed above. For event numbers 54 and 61, it was not possible to constrain the CME-driven shock apex location due to the lack of STEREO imaging, indicated with (NP) in the list. 
 
The coronal shocks usually accelerate at the formation phase, reach their maximum speed between $\sim$3-10 R\textsubscript{$\odot$}, and near $\sim$10-15 R\textsubscript{$\odot$} they start the deceleration phase. Column 7 of Table \ref{Table:SEP_list} shows the maximum speed of the 3D CME-driven shock at the apex, based on the 3D shock reconstruction using a spline fitting to the ellipsoid parameters over time. The uncertainty of the CME-driven shock speed is considered to be 8\% of the value, following \cite{Kwon2014}. In a few events (7 out of 54) the CME-driven shock speed is just smaller than the CME speed but within the uncertainties of the reconstructed parameters. This discrepancy is related to the uncertainty of the fitting process for both the CME and the CME-driven shock; and to the differences in the fitting technique used for the CME-driven shock kinematics estimation, which uses spline fits to the geometrical parameters, as explained by \cite{Kouloumvakos2019} and by \cite{Kouloumvakos2022}.

 \begin{table*}[htb]
\centering
\caption{Spearman and Pearson correlations between the variables involved in this study.}
\label{Table:correlations}
\small
\tabcolsep=0.20cm
\begin{tabularx}{1.0\textwidth}{cHccccHHH} 
\hline
\hline

&Log Peak Intensity& Log Flare Intensity& Log CME-driven shock speed&Log CME speed&Log CME width&delta200\\
&Spearman//Pearson&Spearman//Pearson&Spearman//Pearson&Spearman//Pearson&Spearman//Pearson&Spear/Pear\\
\hline
(1)&()&(2)&(3)&(4)&(5)&(6)\\\hline

 Log SEE Peak Intensity&--&0.32$\pm$0.04//--&0.26$\pm$0.04//--&0.23$\pm$0.03//--&0.21$\pm$0.04//--&--0.09$\pm$0.15//-\\
 (well-connected events)\textsuperscript{*}&(--)&(0.63$\pm$0.03//0.59$\pm$0.03)&(0.53$\pm$0.05//0.65$\pm$0.04)&(0.50$\pm$0.04//0.46$\pm$0.03)&(0.15$\pm$0.06//0.26$\pm$0.04)&(0.02$\pm$0.23//--)&\\\hline
 Log Flare Intensity&-&-&0.50$\pm$0.05//0.53$\pm$0.04 &0.56$\pm$0.03//0.57$\pm$0.03 &0.26$\pm$0.06//0.22$\pm$0.05 &0.13$\pm$ 0.21/-\\\hline
Log CME-driven shock Speed&-&-&-&0.82$\pm$0.02//0.81$\pm$0.02 &0.16$\pm$0.05//0.23$\pm$0.04 &0.07$\pm$0.14/-\\\hline
Log CME Speed&-&-&-&-&0.19$\pm$0.05//0.26$\pm$0.04 &?\\\hline

 \hline
\end{tabularx}
\begin{flushleft}

 \footnotesize{ \textbf{Notes.} * in Col. 1: correlations given for the subsample of well-connected events: -65$^{\circ}$ $\leq$ connection angle $\leq$ 33$^{\circ}$).}

\end{flushleft}
\end{table*}

\begin{table*}[htb]
\centering
\caption{Correlations between the variables involved in this study based on the Kelly method.}
\label{Table:kelly}
\small
\tabcolsep=0.30cm
\begin{tabularx}{0.95\textwidth}{cHccccHHH} 
\hline
\hline

&Log Peak Intensity& Log Flare Intensity& Log CME-driven shock speed&Log CME speed&Log CME width&delta200\\

\hline
(1)&()&(2)&(3)&(4)&(5)&(6)\\\hline
Log SEE Peak Intensity &--& 0.56 $\pm$ 0.38 & 0.68$\pm$0.33 & 0.47 $\pm$ 0.43 & 0.52 $\pm$0.36 \\
 (well-connected events)\textsuperscript{*}&(--)&(0.74 $\pm$ 0.30)&(0.87 $\pm$ 0.20)&(0.60 $\pm$ 0.39)&(0.27 $\pm$ 0.44)&()&\\\hline
 Log Flare Intensity&-&-&0.60 $\pm$ 0.12 & 0.61 $\pm$ 0.11 & 0.25 $\pm$ 0.14 &  \\\hline
Log CME-driven shock speed&-&-&-& 0.88 $\pm$ 0.04 & 0.27 $\pm$ 0.12 & \\\hline
Log CME speed&-&-&-&-&0.28 $\pm$ 0.11 & \\\hline

 \hline
\end{tabularx}
\begin{flushleft}

 \footnotesize{ \textbf{Notes.} * in Col. 1: correlations given for the subsample of well-connected events: -65$^{\circ}$ $\leq$ connection angle $\leq$ 33$^{\circ}$).}

\end{flushleft}
\end{table*}

\section{Relations between SEE parameters and the properties of their parent solar source}
\label{sec:parent_activity_relations}

In this section, we present the relations between the SEE peak intensities and peak-intensity energy spectra and the properties of their parent solar activity for the events measured by MESSENGER. In particular, we compare the SEE events with the X-ray flare characteristics (location, peak intensity), with the 3D kinematic (speed) of the CME and of the CME-driven shock, and with the geometric parameters (width) of the CME, when possible. For this purpose, we use two different probability approaches to apply an appropriate method of correlation between the variables, addressing the instrumental limitations (elevated background level, anti-Sun pointing) of the particle instrument on board MESSENGER.

 \subsection{Frequentist probability approach: Spearman and Pearson correlation coefficients}
 \label{sec:statisitcs_samples}
 There are several methods to approach the correlation between variables, such as Spearman or Pearson techniques. The Spearman rank correlation coefficient \citep[][]{SpearmanAuthor} is often used as a statistical test to determine if there is a relation between two random variables. As a nonparametric rank-based correlation measurement, it can also be used with nominal or ordinal data. The associated statistical test does not need any hypothesis about the shape of the distribution of the population from which the samples are taken \citep{Spearman_Kokoska}. In contrast, the Pearson correlation method  \citep{Kowalski1972Pearson} assumes bivariate normal distribution for the variables. Then, while Pearson correlation provides a complete description of the association when the assumption is fulfilled, conclusions based on significance testing may not be robust in the case of non-bivariate normality. 
Thus, before using the generally known Pearson method in the association of the variables, we characterized the samples to assess whether assumption of normality is acceptable or not. For this purpose, a combination of visual inspection, assessment of the skewness and kurtosis  \citep{west1995structural}, and formal normality tests \citep{D'AgostinoPearson1976normalitytest, Sthephens1974andersontest} were used. We note that, for the variables included in this study, taking logarithms usually transforms a non-Gaussian-like distribution into normality.
 
 Table \ref{Table:statistics_samples} presents a statistics summary of the samples for each of the  parameters of interest, listed in Col. 1. As discussed above, we used the logarithm of the variables in the majority of the parameters to work with normally distributed data. We divided each of the samples into two subsamples (rows): the full sample of events and the sample where the connection angle is -65$^{\circ}$ $\leq$ CA $\leq$ 33$^{\circ}$. This subsample of events is chosen as the well-connected events, as detailed in Sect. \ref{sec:flare_location}. Columns 2-9 show the count, the mean, the standard deviation (STD), the minimum, the 25, 50 (median), 75 percentile marks, and the maximum values, respectively. Column 10 shows the skewness of the sample, which measures the lack of symmetry.  Positive (negative) skewness corresponds to right (left) skewed sample relative to a normal distribution. Column 11 presents the kurtosis value, which measures whether the data are heavy-tailed (kurtosis>3) or light-tailed (kurtosis<3) relative to a normal distribution. Columns 12-13 show the results (stats, p-value) of the Z-tests for the skewness and the kurtosis \citep[e.g.][]{west1995structural}. 
 
 Then, based on the criteria discussed above, we list (Yes) in Col. 14 if the data can be considered normally distributed and (No) if the data show substantial departure from normality, invalidating conventional statistical tests that assume Gaussian distribution (such as, for instance, when estimating the Pearson correlation coefficient). Therefore, in the following we only use the Spearman correlation for the samples with No in Col. 14 of Table \ref{Table:statistics_samples}.  When calculating the correlations, to estimate the statistical uncertainty, namely the confidence intervals of the correlation coefficients derived from the samples, and the uncertainties of the p-value related to the coefficients, we used the Monte Carlo method \citep[e.g.][]{2003WallJenkins,2015Curran}: the correlation coefficient and p-value are calculated for N pairs of values chosen at random within the set of N observations and the respective measurement errors. This procedure is repeated n = 10000 times.    
 
 Furthermore, to characterize the logarithm of the electron peak intensity population in an unbiased fashion, we should address the limitations of the particle instrument on board MESSENGER. The intensity of the SEE events is truncated at the sensitivity
limit (background level) of the EPS instrument, which is close to $\sim$10\textsuperscript{4} (cm\textsuperscript{2} sr s MeV)\textsuperscript{-1}, indicated with the horizontal lines in Fig. \ref{fig:stats_peak_CA}. The truncation indicates that the undetected events are entirely missing from the dataset. This truncation might affect the shape of the distribution of the sample, which can departure from normality and bias the correlation analysis when not properly accounted. Thus, we addressed the truncation characterizing the sample to choose the appropriate correlation method, namely Spearman or Pearson. In addition, due to the anti-Sun pointing of the EPS instrument, MESSENGER observations presumably provide a lower limit to the actual peak intensities of the SEE events. To address this fact, we used the median value of the relation  between the intensities of the antisunward and sunward propagating particles, deduced in Paper I using Solar Orbiter data when the spacecraft radial distance from the Sun ranged from 0.34 to 0.83 au: (I\textsubscript{max\_sun}/I\textsubscript{max\_asun}=1.3$\pm$0.5). Thus, to estimate the correlation coefficient we included the maximum deviation of I\textsubscript{max\_sun}/I\textsubscript{max\_asun}=1.8 in the error of the data points used in the Monte Carlo method discussed above. We note that the SEE peak intensities have asymmetric errors. They may take any value from the measured intensity up to the multiplying factor of 1.8 on the measured intensity. 

\subsection{Bayesian probability approach: Kelly method}
\label{sec:kelly}

We also estimated the correlation coefficients between the different variables included in this study using the Bayesian approach by \cite{Kelly2007}, hereafter the Kelly method. Bayesian inference belongs to the category of evidential probabilities: to evaluate the probability of a hypothesis, the \textit{prior} distribution is specified for each statistical parameter which quantifies the prior knowledge on the possible values. 
This, in turn, is subsequently updated to a \textit{posterior} probability distribution in the light of new, relevant data (evidence). The Bayesian interpretation provides a standard set of procedures and formulae to perform this calculation.
In the Kelly method, a generalized likelihood function for the measured data is constructed and the intrinsic distribution of the independent variables is approximated using a mixture of Gaussian functions instead of using predetermined model distributions. This approach differs from the ones discussed in the previous section and offers a more robust alternative to the commonly used ordinary least-squares (OLS) methods as it directly accounts for: (1) measurement errors in both, the independent and dependent variables in linear regression; (2) intrinsic scatter; and (3) selection effects such as nondetections (e.g. censored or  truncated data) \citep{Kelly2007,2012Feigelson}. 
The uncertainties in the regression parameters and the correlation coefficient are derived from the posterior distribution of the parameters given the observed data \citep{Kelly2007}.
In the following we present the results using both methods, namely the Spearman/Pearson correlations and the Kelly approach.

\begin{figure*}[htb]
\centering
  \resizebox{0.9\hsize}{!}{\includegraphics{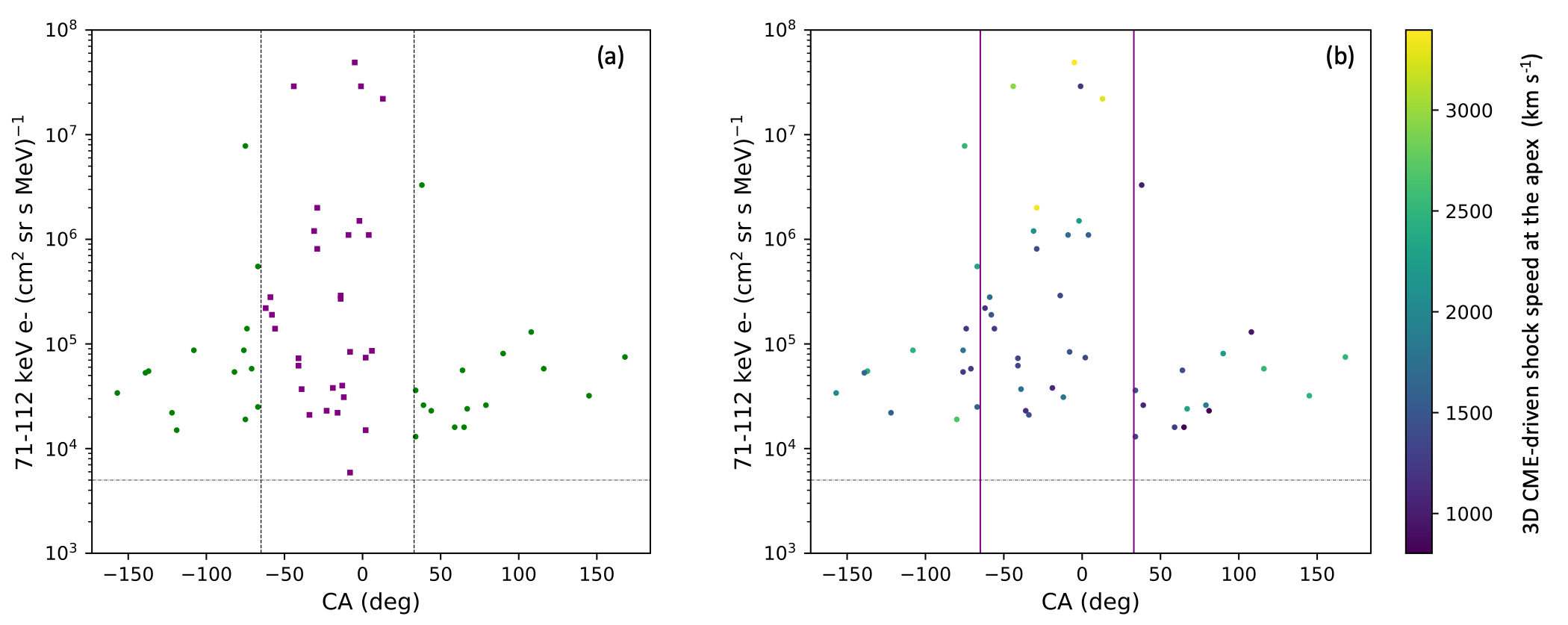}}
     \caption{MESSENGER solar energetic electron peak intensities versus the connection angle (CA). The vertical dashed lines indicate the connection angles CA=-65$^{\circ}$ (left) and CA=+33$^{\circ}$ (right). The horizontal lines show the truncation level of the sample. \textit{(a)} Includes all SEE events selected for the study. The color of the points depends on the CA. The purple squared-shaped points correspond to the sample of well-connected events, namely -65$^{\circ}$ $\leq$ CA $\leq$ +33$^{\circ}$. The rest of the sample is indicated with green circles. \textit{(b)} Only events accompanied by a CME-driven shock are shown, which are color-coded by the shock speed at the apex. }
     \label{fig:stats_peak_CA}
\end{figure*}
    \begin{figure*}
\centering
  \resizebox{0.90\hsize}{!}{\includegraphics{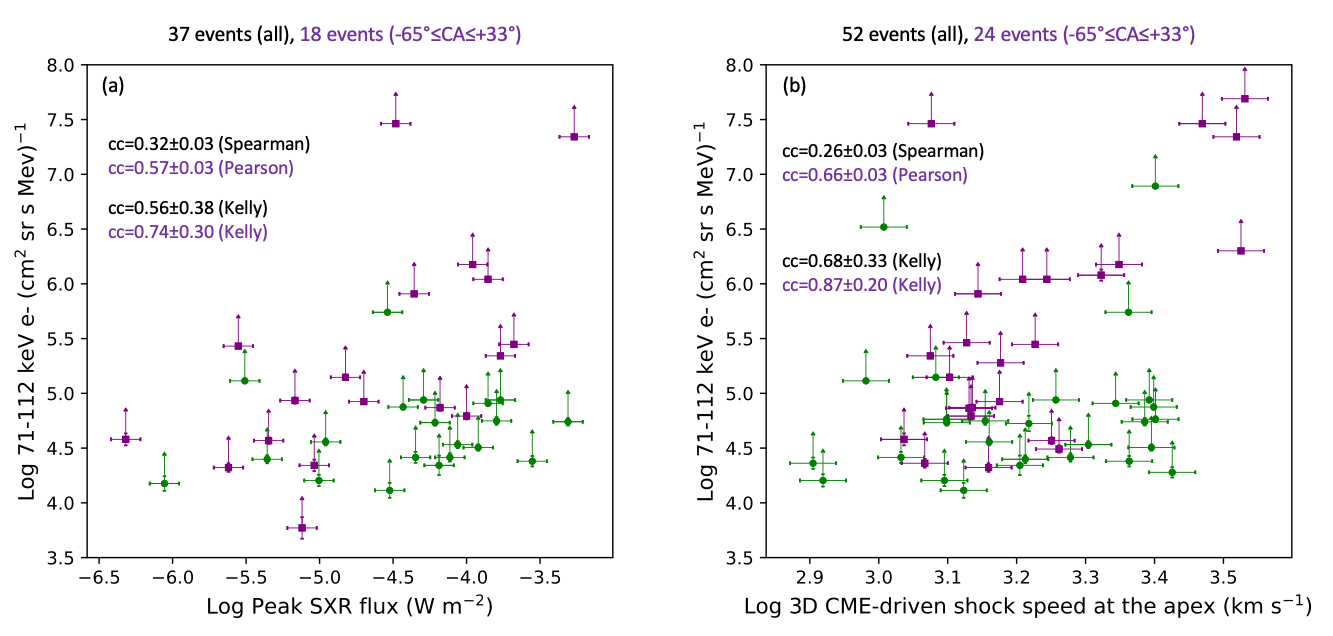}}
     \caption{Logarithm of the electron peak intensity against the logarithm of the flare intensity (\textit{a}) and  the logarithm of the 3D CME-driven shock maximum speed at the apex (\textit{b}). The color code of the points is the same as in Fig. \ref{fig:stats_peak_CA}a. All the points show the error bars corresponding to the uncertainties of the measurements. The vertical arrows over the points represent the error due to the anti-Sun pointing of the EPS instrument. The legend shows the number of events and the correlation coefficients corresponding to the full (well-connected) events in black (purple). Details given in the main text. }
     \label{fig:stats_shock_flare}
\end{figure*} 
  \begin{figure*}[htb]
\centering
  \resizebox{0.9\hsize}{!}{\includegraphics{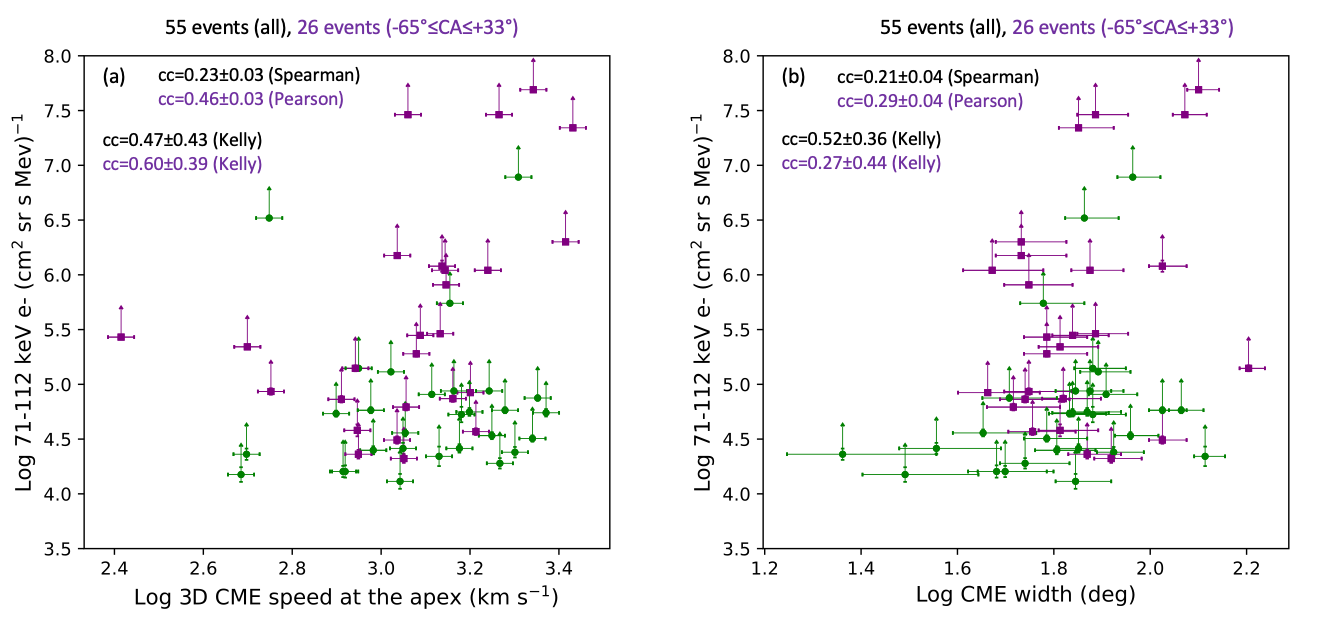}}
     \caption{Logarithm of the electron peak intensity against the logarithm of the 3D CME speed at the apex (\textit{a}) and 3D CME width at the ecliptic plane (\textit{b}).  Colors and legend as in Fig. \ref{fig:stats_shock_flare}.  }
     \label{fig:stats_CME_speed_width}
\end{figure*} 
\begin{figure*}
\centering
  \resizebox{0.9\hsize}{!}{\includegraphics{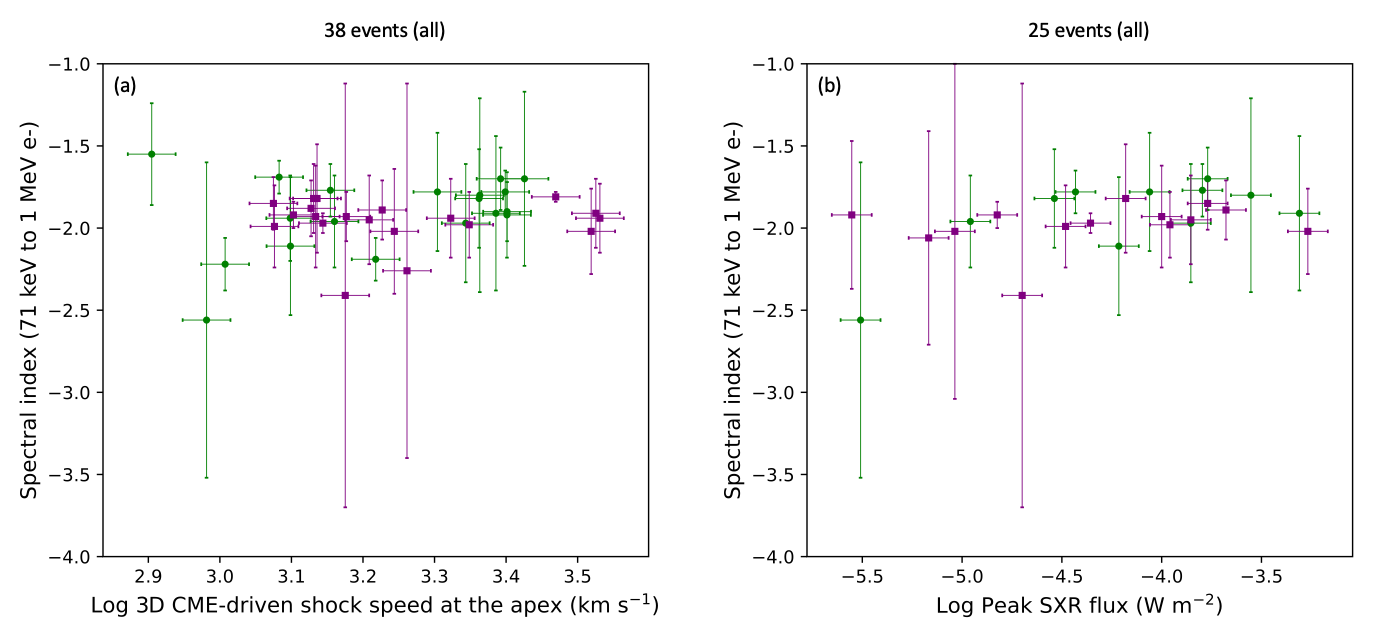}}
     \caption{MESSENGER solar energetic electron spectral indices against the CME-driven shock speed at the apex  \textit{(a)} and the SXR intensity of the flare \textit{(b)}. The color code of the points as in Fig. \ref{fig:stats_peak_CA}a. }
     \label{fig:spectra_flare_shock}
\end{figure*}
\subsection{SEE peak intensity versus solar activity parameters}
\label{sec:peak_solar_activity}

In this section we followed the procedure presented above to estimate the correlations between the SEE peak intensity and the parameters related to their parent solar activity.  The different correlation coefficients found in this study are summarized in Table  \ref{Table:correlations} (Spearman and Pearson methods) and Table \ref{Table:kelly} (Kelly method). Table \ref{Table:correlations} shows that Spearman and Pearson methods give similar results within the uncertainties. Therefore, in the following sections, when both the Spearman and Pearson coefficients are available, we used the Pearson results to compare with previous studies. The correlation coefficients based on the Kelly method listed in Table \ref{Table:kelly} are obtained as the median value from the posterior distribution presented above, while the uncertainty of the correlation corresponds to the median absolute deviation \citep[MAD;][]{2012Feigelson}. 
As detailed below, we note that the values for the correlation coefficients and uncertainties using the Kelly method are larger than the ones obtained using Spearman and Pearson methods. This is mainly due to the Kelly method including the measurement errors when estimating the correlation coefficients. The wider credible intervals are a measure of the widths of the posterior distributions and represent larger uncertainties in the estimated parameters \citep{Kelly2007}. 

\subsubsection{SEE peak intensity versus flare location}
\label{sec:flare_location}

Figure \ref{fig:stats_peak_CA}a shows the 71-112 keV electron peak intensities as a function of the CA, which is the longitudinal separation between the flare location and the footpoint of the magnetic field connecting to the spacecraft, as discussed in Sect. \ref{sec:MESSENGER SEE list}. The events with the largest intensities, between $\sim$10$^{5}$ and $\sim$10$^{8}$ (cm$^{2}$ sr s MeV)$^{-1}$, are observed between -75$^{\circ}\lesssim$ CA$\lesssim$ +38$^{\circ}$, 
including the well-connected events at CA$\sim$0$^{\circ}$, with a trend toward negative CA values. We note the asymmetry in the positive and negative CAs.  We also estimated the connectivity using a range of solar wind speeds of 300-500 km s\textsuperscript{-1}, where the CAs varied between -10$^{\circ}$ and +5$^{\circ}$, not changing the results in the observed asymmetry.  Based on this asymmetry, we divided the full sample into well- and poorly- connected events using the centroid $\phi$\textsubscript{0} and sigma $\sigma$ found by \cite{Lario2013}. They used a Gaussian
to describe the longitudinal distribution
of peak intensities for spacecraft near 1 au. \cite{Lario2013} found, in the case of 71-112 keV electrons and using a constant speed of 400 km s\textsuperscript{-1} for the solar wind speed, that $\phi$\textsubscript{0}=-16$^{\circ}$$\pm$3$^{\circ}$ and $\sigma$=49$^{\circ}$$\pm$2$^{\circ}$. Therefore we chose the connection angle interval CA $\in$ -16$^{\circ}$$\pm$49$^{\circ}$(-65$^{\circ}$ $\leq$ CA $\leq$ +33$^{\circ}$) as the well-connected sector. These SEE events are indicated in purple color in Fig. \ref{fig:stats_peak_CA}a. The green colored events in the figure include the poorly-connected events, which tend to have intensities below $\sim$10$^{5}$ (cm$^{2}$sr s MeV)$^{-1}$. We note that we found a few  higher-intensity events (3 out of 29) in the poorly-connected sample.   The percentage of poorly-connected events included in the full sample is of $\sim$45\%. Figure \ref{fig:stats_peak_CA}b shows that for the majority (25 out 28) of the poorly-connected events the peak electron intensities ranges from $\sim$10$^{4}$ to $\sim$10$^{5}$ (cm$^{2}$sr s MeV)$^{-1}$, independently of the CME-driven shock speed associated to the SEE event. 
 The horizontal lines in Fig. \ref{fig:stats_peak_CA} show the truncation level of the sample, related to the elevated background of the particle instrument, which is close to $\sim$10$^{4}$ (cm$^{2}$sr s MeV)$^{-1}$, as discussed in Sect. \ref{sec:statisitcs_samples}.

\subsubsection{SEE peak intensity versus flare intensity}
 \label{sec:peak_flare_corr}
Figure \ref{fig:stats_shock_flare}a shows the logarithm of the 71-112 keV electron peak intensities plotted against the logarithm of the flare SXR peak flux. The points show the error bars corresponding to the uncertainties of the measurements. The vertical arrows indicate the error in the measured SEE peak intensities due to the anti-Sun pointing of the EPS instrument, as discussed in Sect. \ref{sec:statisitcs_samples}. The color code of the points is the same as in Fig. \ref{fig:stats_peak_CA}a.  The number of events included in the full sample and well-connected events are given on the top of the panel and in Col. 2 of Table \ref{Table:statistics_samples}. The legend also shows the correlation coefficients for the two different approaches discussed above.

Column 2 of the first row in Table \ref{Table:correlations} shows that the Spearman correlation between the logarithms of the peak SEE intensity and the flare intensity is weak: cc=0.32$\pm$0.04.  This correlation is significantly higher than that (cc=0.12) found by \cite{2022Dresing} for a subsample of about 40 electron (55-85 keV) events measured near 1 au by STEREO. We note that that study also included both well- and poorly- connected events. The correlation between the SEE peak intensities and the flare intensity is significantly larger for the well-connected events, with a moderate Pearson correlation coefficient of cc=0.59$\pm$0.03. This value is in agreement with \cite{2015Trottet}, who found a correlation of cc=0.53$\pm$0.09 for the 38 electron (175 keV) events in the western solar hemisphere (CA$\simeq$0), measured near 1 au by the \textit{Advanced Composition Explorer} \citep[ACE;][]{Stone1998ACE}. 

Similarly, using the Kelly approach (first row, Col. 2, Table \ref{Table:kelly}), the logarithm of the peak intensity shows lower correlation with the logarithm of the flare intensity for the full sample (cc=0.56$\pm$0.38) than for the well-connected events (cc=0.74$\pm$0.30). In contrast to Spearman and Pearson methods, the Kelly approach respectively gives a moderate (versus weak) and strong (versus moderate) correlation for both aforementioned samples.

\subsubsection{SEE peak intensity versus CME-driven shock speed}
\label{sec:stats_peak_shock}

 Figure \ref{fig:stats_shock_flare}b shows the logarithm of the 71-112 keV electron peak intensities plotted against the logarithm of the 3D CME-driven shock maximum speed at the apex. Color code of the points as in Fig. \ref{fig:stats_peak_CA}a.  Error bars and legend are similar to Fig. \ref{fig:stats_shock_flare}a. Column 3 of the first row of Table \ref{Table:correlations} shows that the Spearman correlation between the logarithms of the peak SEE intensity and the 3D CME-driven shock speed at the apex is low: cc=0.26$\pm$0.04. This correlation is similar to that (cc=0.24) found by \cite{2022Dresing} for the correlation between the logarithm of the peak intensities and the speed (not the logarithm) of the shock apex for a full sample of 33  electron (55-85 keV) events measured near 1 au by the two STEREO spacecraft. We note that \cite{2022Dresing} include both well- and poorly- connected events and also used 3D parameters of the coronal shock reconstruction, which resulted in larger correlation coefficients compared to not using 3D parameters.

 In the case of well-connected events, indicated in Fig. \ref{fig:stats_peak_CA}b with the vertical purple lines, the correlation found in this study is significantly larger, with a moderate Pearson correlation coefficient of cc=0.65$\pm$0.04. This correlation is slightly higher than that (cc=0.49) estimated  by \cite{Xie2019} comparing the logarithm of the 62-105 keV electron peak intensities with the 3D shock speed (not the logarithm) for CA=0, from a sample of events measured during solar cycle 24 by STEREO and ACE. 

 Similarly, using the Kelly approach (first row, Col. 3, Table \ref{Table:kelly}), the logarithm of the peak intensity shows lower correlation with the logarithm of the shock speed for the full sample (cc=0.68$\pm$0.33) than for the well-connected events (cc=0.87$\pm$0.20). In contrast to Spearman and Pearson methods, the Kelly approach respectively gives a strong (versus weak) and very strong (versus strong) correlation for both aforementioned samples. We note that the uncertainty for the well-connected events is smaller than for the full sample, therefore the significance of the correlation is larger.

\subsubsection{SEE peak intensity versus CME speed}
 \label{sec:peak_CME_corr}

 Figure \ref{fig:stats_CME_speed_width}a shows the logarithm of the 71-112 keV electron peak intensities versus the logarithm of the 3D CME speed at the apex. Color code of the points as in Fig. \ref{fig:stats_peak_CA}a. Error bars and legend are similar to Fig. \ref{fig:stats_shock_flare}. 
 Column 4 of the first row in Table \ref{Table:correlations} shows that the correlation between the logarithms of the peak SEE intensity and the 3D CME speed at the apex is low: cc=0.23$\pm$0.03. This correlation is also significantly larger for the well-connected events, with a moderate Pearson correlation coefficient of cc=0.46$\pm$0.03. This value is slightly lower than that (cc=0.68$\pm$0.09) found by \cite{2015Trottet} for 38 electron (175 keV) events in the western solar hemisphere measured near 1 au by ACE. We note that in that study the values of the CME speed were estimated from linear fits to the time-height trajectory of the CME front, as provided in the CME catalogue \citep{Yashiro2004} of SOHO/LASCO. 

 Similarly, using the Kelly approach (first row, Col. 4, Table \ref{Table:kelly}), the logarithm of the peak intensity shows lower correlation with the logarithm of the CME speed for the full sample (cc=0.47$\pm$0.43) than for the well-connected events (cc=0.60$\pm$0.39). We note that the correlation for the full sample is not significant, as the uncertainty is similar to the coefficient value. In contrast to Spearman and Pearson methods, the Kelly approach gives a strong (versus moderate) correlation for the well-connected sample. We note that in all the cases, the correlations found here are lower than that found for the CME-driven shock presented in Sect. \ref{sec:stats_peak_shock}.

\subsubsection{SEE peak intensity versus CME width}
\label{sec:stats_CME_width}
Figure \ref{fig:stats_CME_speed_width}b shows the logarithm of the SEE peak intensity versus the logarithm of the 3D CME width in the ecliptic plane. Color code of the points as in Fig. \ref{fig:stats_peak_CA}a.  Error bars and legend are similar to Fig. \ref{fig:stats_shock_flare}a.  
Column 5 of the first row in Table \ref{Table:correlations} shows that the correlation between the logarithms of the peak SEE intensity and the 3D CME width is low: cc=0.21$\pm$0.04. This value is in agreement with \cite{1999Kahler}, who found a similar weak correlation (cc=0.28) between the angular width of the CME and the logarithm of the proton peak intensities. The correlation is not significantly larger for well-connected events, with a low Pearson correlation coefficient of cc=0.26$\pm$0.04.

Using the Kelly approach (first row, Col. 5, Table \ref{Table:kelly}), the logarithm of the peak intensity shows higher correlation with the logarithm of the CME width for the full sample (cc=0.52$\pm$0.36) than for the well-connected events (cc=0.27$\pm$0.44). We note that the uncertainty is higher than the correlation coefficient in the case of the well-connected sample, so the correlation is not significant.

\subsection{SEE peak-intensity energy spectra versus coronal shock speed and flare intensity}
In Paper I, the peak-intensity energy spectra of 42 SEE events measured by the MESSENGER mission were analysed. The energies used in the analysis ranged from $\sim$71 keV to $\sim$1 MeV divided into six energy bins. For each of the events, they took the time-of- maximum based on the 71-112 keV channel using one-hour averages and read the intensity peak at this time for the rest of the energy channels. For all the events, the fitting resembled a single power law, giving the spectral index and its uncertainty, $\delta200$. Figures \ref{fig:spectra_flare_shock}a and \ref{fig:spectra_flare_shock}b show the spectral indices against the logarithm of the 3D CME-driven shock maximum speed at the apex and the logarithm of the flare intensity, respectively. The color code of the points is the same as in Fig. \ref{fig:stats_peak_CA}a.  We find no correlation for any of them. This result is in agreement with \cite{2022Dresing}, where no clear correlations are found between the shock parameters and the spectral indices of the near-relativistic electrons. We do not observe harder spectra with increasing peak intensities (not shown), as \cite{2001Kahler} found for high energy protons (>10 MeV). 
 
\subsection{Relations between the solar parent activity}
\label{sec:relations_parent_activity}
The last three rows of Table \ref{Table:correlations} and Table \ref{Table:kelly} show the correlations between the variables describing the solar activity. There is a moderate correlation between the logarithms of the flare intensity and the 3D CME-driven shock speed (Pearson: cc=0.53$\pm$0.04; Kelly: cc=0.60$\pm$0.12; 33 events) and the 3D CME speed (Pearson: cc=0.57$\pm$0.03, Kelly: cc=0.61$\pm$0.11; 36 events). This last value is similar to the correlation coefficient (Pearson: cc=0.66) obtained by \cite{Kihara2020} sampling 79 events from 2006 to 2014 with near 1 au spacecraft. We note that \cite{Kihara2020} use the CME speed instead of the logarithm of this variable. The correlation between the 3D CME-driven maximum shock speed and the 3D CME speed is strong (Pearson: cc=0.81$\pm$0.02; Kelly: cc=0.88$\pm$0.04; 52 events). There is a weak correlation between the 3D CME width and any of the other variables. 

\section{Summary and discussion}
\label{sec:summary_discussion}

We used the list of 61 SEE events measured by the MESSENGER mission from 2010 to 2015, presented by \cite{Rodriguez-Garcia2023_messenger} (Paper I), when the heliocentric distance of the spacecraft varied from 0.31 au to 0.47 au. Due to the elevated background intensity level of the particle instrument on board MESSENGER, the SEE events measured by this mission are necessarily large and intense; most of them are accompanied by a CME-driven shock, are widespread in heliolongitude, and display relativistic ($\sim$1 MeV) electron intensity enhancements. The largest peak intensities, between $\sim$10$^{5}$ and $\sim$10$^{8}$ (cm$^{2}$ sr s MeV)$^{-1}$,  are observed in the range of connection angles -75$^{\circ}<$ CA $<$ +38$^{\circ}$, with an asymmetry to longitudes east of the well-connected longitudes CA $\sim$ 0$^{\circ}$.

To relate the near-relativistic electron peak intensity and the peak-intensity energy spectra to different parameters of the parent solar source, we (1) considered the flare peak intensity measured in Paper I from the events originating on the visible side of the Sun from Earth's point of view and (2) took advantage of the multi-viewpoint spacecraft observations to reconstruct, when possible, the 3D large-scale structure of the CME and the CME-driven shock using the GCS model \citep{Thernisien2011} and the ellipsoid model \citep{Kwon2014}, respectively. We added some of the reconstructed parameters to the list of SEE events in Table \ref{Table:SEP_list} for future reference.

\subsection{Summary of observational results}
In this work, we first characterized the distribution of the samples in order to select the appropriate method to estimate the correlation coefficients between the SEE peak intensity and energy spectra and several parameters related to the solar activity. We also addressed the fact that the peak intensities measured by MESSENGER were truncated and presented asymmetric uncertainties due to instrumental limitations (elevated background level, anti-Sun pointing).  The observational results of this study are summarised as follows:

\begin{itemize}
    \item[$\bullet$] There is an asymmetry in the positive and negative connection angles for which the largest intensities are measured. This asymmetry was, therefore, considered in the definition of the connection angle interval in the so-called well-connected events, namely -65$^{\circ}\leq$ CA $\leq+33^{\circ}$, in which we found stronger correlations between the SEE peak intensities and the solar parameters in comparison to the full sample.

    \item[$\bullet$] In the majority of the poorly-connected events, the peak intensities are below $\sim$10\textsuperscript{5} (cm\textsuperscript{2} sr s MeV)\textsuperscript{-1}, independently of the intensity of the flare or the speed of the CME-driven shock. A poor connectivity to the source weakens the correlations between the peak intensities and the different parent solar source parameters. 
    
    \item[$\bullet$] The strongest correlations are found between the near-relativistic electron peak intensities and the maximum speed at the apex of the 3D CME-driven shock and the flare intensity for the so-called well-connected events. The Pearson correlation coefficients and their uncertainties based on a Monte Carlo method are: cc=0.65$\pm$0.04 (shock) and cc=0.59$\pm$0.03 (flare).  We note the similar correlations between the SEE peak intensities and the 3D CME-driven shock maximum speed at the apex  and the flare intensity. The correlation coefficients based on the Kelly method are cc=0.87$\pm$0.20 (shock) and cc=0.74$\pm$0.30 (flare). We also note the reduced uncertainty in the case of the shock sample compared to the correlations with other solar parameters.

    \item[$\bullet$] We find a moderate correlation between the near-relativistic electron peak intensities and the CME speed at the apex for the well-connected events,  with a Pearson (Kelly) correlation coefficient of cc=0.46$\pm$0.03 (cc=0.60$\pm$0.39).
    
    \item[$\bullet$] The weakest correlations for the well-connected events are found between the near-relativistic electron peak intensities and the 3D CME width, with a Pearson (Kelly) correlation coefficient of cc=0.26$\pm$0.04 (cc=0.27$\pm$0.44). We note that in the Kelly method, the uncertainty is higher than the correlation coefficient,  indicating that no significant correlation can be determined. 
    
    \item[$\bullet$] The correlations between the near-relativistic electron peak intensities and the solar activity, namely the flare intensity and 3D CME-driven shock speed, estimated in this study are higher than that found by two equivalent studies  based on near 1 au measurements, namely, \cite{2022Dresing} and \cite{Xie2019}. We note, however, that in the study by \cite{2015Trottet}, the correlations with the flare intensity are similar.

    \item[$\bullet$] We find no correlation between the spectral indices and the flare intensity nor the CME-driven shock speed.

    \item[$\bullet$] Correlations of similar order exist between the different parameters describing the solar activity, such as the flare intensity, the CME speed, or the CME-driven shock.  The correlation between the solar activity parameters (e.g. flare intensity and shock speed) is smaller than the correlations between the SEE peak intensities and either the flare intensity or the shock speed.

\end{itemize}

\subsection{Effect on the interpretation of the origin of SEEs}
\subsubsection{Correlations between the parameters characterising the solar activity}

One of the difficulties found when interpreting statistical relations between solar activity and SEEs is the interrelationship of the different parameters utilized to characterise the solar activity, as summarized in the last three rows of Tables \ref{Table:correlations} and \ref{Table:kelly}. For example, \cite{1982Kahler} introduced the term Big Flare Syndrome to illustrate the observational fact that there is a correlation between any two parameters measuring the magnitude of a flare event, independent of the detailed physical relationship between them. In this study, we find a moderate correlation between the SXR peak flux and the CME speed (Pearson: cc=0.57$\pm$0.03; Kelly: cc=0.61$\pm$0.11). This correlation might be related to a common physical process at the Sun. It is well known that the acceleration of CMEs is closely related in time with the evolution of thermal energy release in the associated flare \citep{2004Zhang,2012Bein}, suggesting an interdependence between the CME speed and the peak flux of the flare.

As expected, we find a strong correlation between the maximum speed at the apex of the 3D CME-driven shock and the 3D CME speed at the apex derived from a linear fit of the time evolution of the CME apex height (Pearson: cc=0.81$\pm$0.02; Kelly: cc=0.88$\pm$0.04). This correlation might be expected to be even higher (i.e., closer to cc=1). A reason for the rather lower value we find could be related to measuring the maximum speed for the shock but the average linear speed for the CME. Lastly, we find a moderate correlation between the flare intensity and the 3D CME-driven shock speed (Pearson: cc=0.53$\pm$0.04; Kelly: cc=0.60$\pm$0.12), which might be related to both the intrinsic relation between the flare intensity and the CME speed, and the relation between the CME speed and the CME-driven shock speed, as discussed above. 

Using partial correlations in the analysis of the relations between SEE parameters and the solar activity  \citep[e.g.][]{2015Trottet} might be a simplification of the real picture, as the correlations between variables actually show a degeneracy in the parameter space. The flare-related and CME-related phenomena are expressions of the solar activity and it is very probable that both of them share the same common origin at the Sun. Furthermore, based on previous studies, it is probable that using different parameters to characterize the solar activity, such as the fluence for the flare activity \citep{2015Trottet} or the CME-driven shock speed at the cobpoint \citep[][]{1995Heras} for the shock activity \citep{2022Dresing}, would increase the correlations with the peak intensities.

\subsubsection{Correlation between SEE intensities and solar activity}
\label{subsec:corrl_solar}

Our study finds a distinct difference between the SEE correlations found for different samples, when classifying them by different connection angles. For the full sample of the events, including poorly connected events, we find similar weak Pearson correlations for the different quantities that describe the solar activity, varying from cc $\sim$0.21 to $\sim$0.32. Also, the uncertainty found in the correlations based on the Kelly method are significant with respect to the correlation coefficient, meaning that the correlations are not clear. This results is expected due to the inclusion of the poorly-connected events in the study, where the transport effects and/or the connection to peripheral areas of the source (shock) could significantly distort the correlations. This behaviour is clearly observed in Fig. \ref{fig:stats_peak_CA}b, where the majority of the points outside the purple vertical lines, indicating the well-connected range, are showing intensities between $\sim$10\textsuperscript{4} and  $\sim$10\textsuperscript{5}(cm\textsuperscript{2} sr s MeV)\textsuperscript{-1} independently of the shock speed. The few high-intensity points outside the well-connected range might be related to varying CME widths and/or different footpoint locations caused by non-nominal solar wind speed or disturbed Parker-field.

However, for the well-connected events, namely for -65$^{\circ}$ $\leq$ CA $\leq$ 33$^{\circ}$, we generally find clearer correlations. The SEE peak intensities correlate similarly with the 3D CME-driven maximum shock speed (Pearson: cc=0.65$\pm$0.04; Kelly: cc=0.87$\pm$0.20) and with the SXR peak flux (Pearson: cc=0.59$\pm$0.03; Kelly: cc=0.74$\pm$0.30). We note the lower uncertainties found in the Kelly method between the SEE peak intensities and the shock speed in comparison with the flare intensity. However, the samples are not the same, as we are not considering far-side events for the flare intensity and this fact could affect the comparison between correlations. These correlations are   better than with the CME speed (Pearson: cc=0.46$\pm$0.03; Kelly: cc=0.60$\pm$0.39).  Thus, the correlation of the peak electron intensity is  higher and also more significant with the 3D CME-driven shock maximum speed at the apex than with the CME speed at the apex. This means that the maximum shock speed might be a better proxy of the acceleration of energetic electrons than the linear CME speed. 

 For the well-connected events, we also found that the correlation between the logarithms of the peak intensity of the SEE events and the speed of the CME-driven shock at the apex is stronger in the SEE events measured by MESSENGER (Pearson: cc=0.65$\pm$0.04; Kelly: cc=0.87$\pm$0.20) in comparison to near 1 au data (Pearson: cc=0.49) for similar near-relativistic energies and using CME and associated 3D shock parameters \citep{Xie2019}. We note that \cite{Xie2019} use the speed of the shock instead of the logarithm of the shock speed. However, the results are similar comparing directly the shock speed in both studies. 
 Similarly, the correlation between the peak intensities and the flare intensity for the full sample (Pearson: cc=0.32$\pm$0.04; Kelly: cc=0.56$\pm$0.38) is higher than that found for near 1 au measurements \citep[Pearson: cc=0.12;][]{2022Dresing}, which included both well- and poorly- connected events. In the case of well-connected events (-65$^{\circ}$ $\leq$ CA $\leq$ 33$^{\circ}$), we find similar correlations (Pearson: cc=0.59$\pm$0.03; Kelly: cc=0.74$\pm$0.30) as \cite{2015Trottet}, using near 1 au data (Pearson: cc=0.53$\pm$0.09). 
 
On a statistical basis the CME width seems not to play a relevant role in terms of the peak intensity of the SEE event, as the correlations both for the full sample and well-connected events are weak. We note that we found slightly higher correlations between the peak intensities and the CME width estimated as in \cite{Dumbovic2019} that takes the tilt of the CME into account (Pearson: cc=0.21$\pm$0.04 for the full sample, cc=0.26$\pm$0.04 for the well-connected events), than the face-on width of the CME (Pearson: cc=0.11$\pm$0.05, cc=0.16$\pm$0.05, not shown). In the case of the spectral indices, on top of the large uncertainties, we suspect that the missing correlations might be partly due to a selection effect, as MESSENGER is mostly measuring large events, the majority of them being widely spread in the heliosphere and with the presence of relativistic electron enhancements.  The spectral indices in the MESSENGER sample are mainly hard, with a mean $\delta$200=-1.9$\pm$0.3, as can be observed in Fig. \ref{fig:spectra_flare_shock}, while SEE spectra in general can be much softer \citep[e.g.][]{Dresing2020}.

 \subsubsection{Other quantities affecting the peak intensities}

  The conditions of particle acceleration and propagation in the high corona and interplanetary space affect SEP intensities. The pre-event intensity level might also play a role. Figure \ref{fig:stats_shock_flare}b shows that, for the well-connected events (purple points), the peak SEE intensities associated to a CME-driven shock of a given speed vary over $\sim$4 orders of magnitude, similar to the result found by \cite{2001Kahler}, who used pre-event background-subtracted SEE peak intensities. This could be interpreted as evidence for a supra-thermal seed population that
made local shock acceleration more efficient. Other related factors to observing a range of peak intensities for a given speed might be the dynamic connection between MESSENGER and the traveling shock; and the presence of previous disturbances in the IP space that may affect the interplanetary magnetic-field structure in which SEEs propagate.
 
   The asymmetry in the positive and negative angles delimiting the subsample of events with the highest peak intensities might be associated to several processes. A possible scenario could be related to acceleration mechanisms in the shock environment at a certain height from the Sun and the evolution of magnetic field connection to the shock front \citep[e.g.][]{2014Larioasymmetry, Ding2022}, where the maximum peak intensity is observed when the flare occurs eastward of the spacecraft magnetic footpoint. For example, the nominal best connection for an observer near 0.4 au, using a speed of 400 km s\textsuperscript{-1}, to a source at W30 is modified by a CME-driven shock that moves out radially so that the connection to its apex is more towards the east than W30. Perpendicular diffusion processes during the transport of SEPs in the heliosphere might also be related to this asymmetry \citep[e.g.][]{2015He}.

As MESSENGER lacks solar wind measurements, the magnetic separation angle, determined with an assumed solar wind speed of 400 km s\textsuperscript{-1}, could deviate significantly. However, as discussed above, we obtained similar results regarding the observed asymmetry when using a range of solar wind speeds of 300-500 km s\textsuperscript{-1}. We also clearly observe in this study that the poor connectivity to the solar source blurs the correlation between the peak intensities and the solar activity. This might be related to  the poorly-connected events being affected by transport effects and/or to the connection to weaker parts of a shock, as discussed in Sect. \ref{subsec:corrl_solar}. 

\subsection{Final discussion}

The highest correlations found in this study between the near-relativistic electron peak intensities and the solar activity are with the 3D CME-driven shock speed and the flare intensity. This is a statistical confirmation of the idea that both flare and shock-related processes may contribute to the acceleration of near relativistic electrons in large SEE events \citep{Kallenrode2003, 2015Trottet,2022Dresing}, provided the flare-accelerated particles escape to
interplanetary space.  The correlations found between the flare intensity and the shock speed being lower than the correlations between the SEE peak intensities and the flare intensity or the shock speed might support this result. 

Also, we found a stronger correlation between the SEE peak intensities and the maximum speed of the 3D CME-driven shock than with the 3D CME speed. This means that the maximum speed of the CME-driven shock, usually observed below 10 R\textsubscript{$\odot$}, is a better proxy to investigate particle acceleration related mechanisms than the CME speed from linear fits to the height-time profile in the coronagraph field-of-view, as usually used in the past \citep[e.g.][]{2015Trottet, Kihara2020}.

Closer to the Sun (i.e., closer to the acceleration site), we find stronger correlations with the solar parameters associated to the electron acceleration mechanisms in comparison with some of the previous studies using near 1 au data. This difference is more relevant when comparing studies with similar connectivity and using the 3D parameters of the CME-related activity \citep{Xie2019, 2022Dresing}. This suggests that the effect of the IP transport from near 0.3 au to near 1 au on the energetic electrons might weaken the correlations between the solar source parameters and the peak intensities measured in situ. However, the correlations found by \cite{2015Trottet} are similar as that found in this study. Then, future studies with same samples and following the same methodology near 0.3 au and near 1 au are necessary to investigate this possible effect of the IP transport further. 

Two interesting observational results of this study are (1) the asymmetry to the east of the range of connection angles for which the SEE events present the highest peak intensities, and (2) the presence of relativistic electrons in 37 out of 61 SEE events. Previous studies related these observations to different acceleration mechanisms. In the case of the presence of MeV electrons, both flare-related \citep[e.g.][]{1974Simnett} and shock-related \citep[e.g.][]{2007Kahler, 2022Dresing} scenarios have been suggested. In the case of the asymmetry to the east, it has been attributed to diverse factors such as acceleration mechanisms in the shock environment \citep[e.g.][]{2014Larioasymmetry, Ding2022} or the role of perpendicular diffusion in the particle transport \citep[e.g.][]{2015He}, as discussed above. Based on the comparison of the correlation coefficients presented in this study alone, we do not find statistical indication to favour one mechanism over another (i.e. the differences in the correlation coefficients are not statistically significant).  Therefore, the analysis and outcomes presented here might be further investigated with data from the new ongoing missions exploring the innermost regions of the heliosphere, such as \textit{Solar Orbiter} \citep[][]{Muller2020,Zouganelis2020}, \textit{Parker Solar Probe} \citep[PSP;][]{Fox2016} and \textit{BepiColombo} \citep[][]{2010Benkhoff}, together with near 1 au missions remotely observing the Sun. Alternative parameters related to the solar activity, such as the SXR fluence, the CME over expansion speed in the early phases close to the Sun surface, and shock characteristics at the cobpoint, which might describe in some sense better the strength of the probable accelerators \citep{2015Trottet, 2022Dresing}, could be also investigated in future studies. By using these new multi-spacecraft observations and as we progress into the solar cycle 25, we will measure more events and increase the statistics, which will allow a reduction of the uncertainties.

\begin{acknowledgements}
 The UAH team acknowledges the financial support by the Spanish Ministerio de Ciencia, Innovación y Universidades FEDER/MCIU/AEI Projects ESP2017-88436-R and PID2019-104863RB-I00/AEI/10.13039/501100011033 and by the European Union’s Horizon 2020 research and innovation program under grant agreement No. 101004159 (SERPENTINE). LRG is also supported by the European Space Agency, under the ESA/NPI program and thanks Karl Ludwig Klein, Angelos Vourlidas, and Nariaki Nitta for their help. LAB acknowledges the support from the NASA program NNH17ZDA001N-LWS (Awards Nr. 80NSSC19K0069 and 80NSSC19K1235). AK acknowledges financial support from NASA NNN06AA01C (SO-SIS Phase-E) contract. ND is grateful for support by the Turku Collegium for Science, Medicine and Technology of the University of Turku, Finland and acknowledges the support of Academy of Finland (SHOCKSEE, grant 346902). DL acknowledges support from NASA Living With a Star (LWS) programs NNH17ZDA001N-LWS and NNH19ZDA001N-LWS, and the Goddard Space Flight Center Internal Scientist Funding Model (competitive work package) program and the Heliophysics Innovation Fund (HIF) program.  The authors acknowledge the different SOHO, STEREO instrument teams, and the STEREO and ACE science centers for providing the data used in this paper. This research has used PyThea v0.7.3, an open-source and free Python package to reconstruct the 3D structure of CMEs and shock waves (GCS and ellipsoid model). 

\end{acknowledgements}

\begin{flushleft}

\textbf{ORCID iDs} 
\vspace{2mm}

Laura Rodríguez-García \orcid{https://orcid.org/0000-0003-2361-5510}

Laura Balmaceda \orcid{https://orcid.org/0000-0003-1162-5498}

Raúl Gómez-Herrero \orcid{https://orcid.org/0000-0002-5705-9236}

Athanasios Kouloumvakos \orcid{https://orcid.org/0000-0001-6589-4509}

Nina Dresing \orcid{https://orcid.org/0000-0003-3903-4649}

David Lario
\orcid{https://orcid.org/0000-0002-3176-8704}

Ioannis Zouganelis
\orcid{https://orcid.org/0000-0003-2672-9249}

Annamaria Fedeli \orcid{https://orcid.org/0000-0001-9449-4782}

Francisco Espinosa Lara \orcid{https://orcid.org/0000-0001-9039-8822}

Ignacio Cernuda
\orcid{https://orcid.org/0000-0001-8432-5379}

George Ho
\orcid{https://orcid.org/0000-0003-1093-2066}

Robert Wimmer-Schweingruber \orcid{https://orcid.org/0000-0002-7388-173X}

Javier Rodríguez-Pacheco
\orcid{https://orcid.org/0000-0002-4240-1115}

\end{flushleft}
%
%
\bibliographystyle{bibtex/aa}
\bibliography{bibtex/biblio.bib}

\onecolumn
\begin{appendix}

\section{Solar energetic electron events measured by the MESSENGER mission} 
\label{appendix:MESSENGER_table}
%

\onecolumn
\fontsize{5}{7}\selectfont
\begin{ThreePartTable}
\begin{TableNotes}

\item \footnotesize{\textbf{Notes. }Columns 1 and 2: Event number and date. Column 3: Type III radio burst onset time. Column 4: Flare location in Stonyhurst coordinates and flare class based on GOES Soft X-ray (SXR) peak flux. Column 5: 3D CME speed at the apex based on the GCS analysis. Column 6: 3D CME width at the equatorial plane based on the GCS reconstructed CME parameters, as in  \cite{Dumbovic2019}. Column 7: 3D CME-driven shock maximum speed at the apex based on the ellipsoid model \citep{Kwon2014}. Column 8: Longitudinal separation between the flare location and the footpoint of the magnetic field line connecting to MESSENGER, based on a 400 km s$^{\ -1}$ Parker spiral (positive connection angle (CA) denotes a flare source located at the western side of the spacecraft magnetic footpoint). Column 9: MESSENGER radial distance from the Sun. Column 10: 71 -112 keV  electron peak intensity measured by MESSENGER. The pre-event background level is shown in parenthesis. Column 11: Spectral index of peak intensities based on 71 keV to 1 MeV energies.  * in Col. 1: Widespread SEP event, namely when MESSENGER |CA| or |CA difference| with near 1 au spacecraft is $\geq$80$^{\circ}$.    \textsuperscript{ˆ} in Col. 3: Type III radio burst onset time is uncertain due to occultation or multiple radio emission at the same time during the onset of the event. \textsuperscript{§} in Col. 4: The GOES intensity level is deduced from the STEREO/EUVI light curve as explained in \cite{2021Rodriguez-Garcia}.  NP in Cols. 5-7: not possible to reconstruct. \textsuperscript{!} in Cols. 5-7: CME and CME-driven shock reconstructions using only LASCO and SDO data.  \textsuperscript{$\dagger$} in Col. 11: Presence of $\sim$1 MeV electrons. }

\end{TableNotes}

\setlength{\tabcolsep}{5pt}
\renewcommand{\arraystretch}{0.8}
\fontsize{7}{9}\selectfont
\begin{longtable}{cccccccccccHH}

\caption{Solar energetic electron events measured by MESSENGER.  }
\label{Table:SEP_list}
\\
\hline
\hline
\multicolumn{4}{c}{Solar event}&\multicolumn{2}{c}{CME parameters}&Shock&&& \multicolumn{3}{c}{SEE event}&Lists\\
\cline{2-3} \cline{5-6} \cline{10-12}
\#&Date &T-III &Flare & speed &width&speed&CA &R& I\textsubscript{max\_MESS} (bg)&$\delta$&I\textsubscript{max\_near\_1au}[est.](bg) (s/c: CA diff.) \\
& & onset& loc [class]&\multicolumn{2}{c}{(GCS)}&(3D)& MESS & MESS & 71 to 112 keV e&MESS&75 to 105 keV e \\
& &(UT $\pm$ 5 min)&(deg)&(km s$^{-1}$)&(deg)&(km s$^{-1}$)&(deg)&(au)&(cm\textsuperscript{2} sr s MeV)\textsuperscript{-1}&(-)&(cm\textsuperscript{2} sr s MeV)\textsuperscript{-1}\\
 \hline
(1)&(2)&(3)&(4)&(5)&(6)&(7)&(8)&(9)&(10)&(11)&(12)&(13)\\
 \hline
 \endfirsthead
 
\caption{(continued.)}\\
\hline
\hline
\multicolumn{4}{c}{Solar event}&\multicolumn{2}{c}{CME parameters}&Shock&&& \multicolumn{3}{c}{SEE event}&Lists\\
\cline{2-3} \cline{5-6} \cline{10-12}
\#&Date &T-III &Flare & speed &width&speed&CA &R& I\textsubscript{max\_MESS} (bg)&$\delta$&I\textsubscript{max\_near\_1au}[est.](bg) (s/c: CA diff.) \\
& & onset& loc [class]&\multicolumn{2}{c}{(GCS)}&(3D)& MESS & MESS & 71 to 112 keV e&MESS&75 to 105 keV e \\
& &(UT $\pm$ 5 min)&(deg)&(km s$^{-1}$)&(deg)&(km s$^{-1}$)&(deg)&(au)&(cm\textsuperscript{2} sr s MeV)\textsuperscript{-1}&(-)&(cm\textsuperscript{2} sr s MeV)\textsuperscript{-1}\\
 \hline
(1)&(2)&(3)&(4)&(5)&(6)&(7)&(8)&(9)&(10)&(11)&(12)&(13)\\
 \hline
 \endhead
\multicolumn{13}{c}{(Continued on next page.)}

\endfoot

\insertTableNotes  
\endlastfoot

*1 &2010/08/14 &10:00\textsuperscript{ˆ}&N17W052 [C4.4]&960&64&1631&-67 &0.31&2.5$\times$10\textsuperscript{4} (1.6$\times$10\textsuperscript{4})&-& 7.5e1 [1.2e2] (4.0e1) (STA: -20$^{\circ}$)&a, b, c, f, g \\
*2 &2010/08/18 &05:35&N17W101 [C4.5]&1634&57&1781&-39 &0.31 &3.7$\times$10\textsuperscript{4} (1.5$\times$10\textsuperscript{4}) &-& 3.2e3 [3.1e3] (3.2e1) (STA: +1$^{\circ}$)&a, b, c, g \\
*3&2011/03/07 &19:55\textsuperscript{ˆ}&N30W048 [M3.7]&2250&51&2505&168  &0.34  & 7.5$\times$10\textsuperscript{4} (1.6$\times$10\textsuperscript{4})  & -1.78$\pm$0.13\textsuperscript{$\dagger$}&-&b, c, d, e, g \\
*4 &2011/06/04 &06:50&N16W144 [-]&1086&106&1826& -12 &0.33  &3.1$\times$10\textsuperscript{4} (9.0$\times$10\textsuperscript{3}) &-2.26$\pm$1.14 &7.5e3 [6.8e3] (7.9e1) (STA: +3$^{\circ}$)&a, b, g \\
*5&2011/06/04 &21:50\textsuperscript{ˆ}&N16W153 [-]&2200&126&3397&-5  &0.33  & 4.9$\times$10\textsuperscript{7} (2.0$\times$10\textsuperscript{4})  & -1.94$\pm$0.21\textsuperscript{$\dagger$} &6.3$\times$10\textsuperscript{5} [5.9$\times$10\textsuperscript{5}] (3.0e4) (STA: +5$^{\circ}$)&a, b, g \\
*6&2011/08/02 &06:25\textsuperscript{ˆ}&N15W015 [M1.4]&807&90&1114& 19 &0.46&1.5$\times$10\textsuperscript{3} (2.5$\times$10\textsuperscript{2})&-& 1.0e2 [4.6e2] (2.3e1) (STB: +24$^{\circ}$)&a, b  \\
*7 &2011/08/04 &03:50&N19W036 [M9.3]&1125&88&2572& 37 &0.46&1.6$\times$10\textsuperscript{3}(5.0$\times$10\textsuperscript{2})  &- & 2.3e2 [7.5e2] (3.3e1) (STB: +27$^{\circ}$)&a, b, c, d, e, f, g  \\
*8&2011/09/22 &10:40&N09E089 [X1.4]&1300&81&2206& 90 &0.36&8.1$\times$10\textsuperscript{4} (1.4$\times$10\textsuperscript{4})  &-1.97 $\pm$0.36\textsuperscript{$\dagger$}& 4.5e3 [6.4e3] (9.4e2) (STA: +18$^{\circ}$)& a, b, c, d, e, f, g   \\
*9&2011/10/04 &12:30\textsuperscript{ˆ}&N26E153 [-]&1358&77&1341& -14 &0.42&2.9$\times$10\textsuperscript{5} (1.0$\times$10\textsuperscript{4})  &-1.88$\pm$0.17\textsuperscript{$\dagger$} &-&a, b, c, e, g  \\
10&2011/10/14 &11:00\textsuperscript{ˆ}&N10E140 [-]&889&74&1166& -23 &0.47&2.3$\times$10\textsuperscript{4} (1.2$\times$10\textsuperscript{4})  &-&-&a, b\\
*11\ &2011/11/03 &22:15&N09E154 [-]&890&76&1210& -74 &0.44&1.4$\times$10\textsuperscript{5} (9.0$\times$10\textsuperscript{3})  &-1.69$\pm$0.10\textsuperscript{$\dagger$}&-&a, b, c, d, e, f, g  \\
12 &2011/11/09 &13:10&N24E035 [M1.1]&1133&45&1446& 34 &0.42&3.6$\times$10\textsuperscript{4} (1.0$\times$10\textsuperscript{4}) &-1.96$\pm$0.28\textsuperscript{$\dagger$} &9.1e2 [9.7e1] (9.2e1) (STB: -32$^{\circ}$)&a, b, g \\
*13 &2011/11/17 &20:15\textsuperscript{ˆ}&N18E120 [-]&948&106&1254&-71 &0.38&5.8$\times$10\textsuperscript{4} (7.1$\times$10\textsuperscript{3})  &-1.94$\pm$0.26\textsuperscript{$\dagger$}& 7.9e2 [1.0e3] (2.7e2) (STB: -11$^{\circ}$)&a, g \\
*14&2012/01/02&14:30&N08W104 [C2.4]&1125&83&1443&-34&0.43&2.1$\times$10\textsuperscript{4} (8.1$\times$10\textsuperscript{3})&-& 6.4e2 [1.7e3] (3.7e2) (STA: -29$^{\circ}$)& b, g \\
*15&2012/01/23&03:40&N28W021 [M8.7]&1775&91&2014&-157&0.46&3.4$\times$10\textsuperscript{4} (8.7$\times$10\textsuperscript{3})& -1.78$\pm$0.36\textsuperscript{$\dagger$}&  1.8e4 [1.6e4] (7.8e1) (STA: +11$^{\circ}$)& b, c, d, e, f, g\\
*16 &2012/01/27 &18:15&N27W078 [X1.7]&1750&70&2468& -108 &0.46&8.7$\times$10\textsuperscript{4} (8.5$\times$10\textsuperscript{3})  &-1.70$\pm$0.19\textsuperscript{$\dagger$}& 1.6e4 [1.1e4] (1.0e4)
(STA: +19$^{\circ}$)&a, b, c, d, e, f, g  \\
*17&2012/03/04 &11:05&N19E061 [M2.0]&1588&46&1497& -8 &0.31&8.4$\times$10\textsuperscript{4} (8.9$\times$10\textsuperscript{3})  &-2.41$\pm$1.29\textsuperscript{$\dagger$}&  5.5e4 [5.3e4] (6.2e2) (STB: +1$^{\circ}$) & b, c, f, g \\
*18 &2012/03/05 &03:35&N17E052 [X1.1]&850&72&2231& -2 &0.31&1.5$\times$10\textsuperscript{6} (4.1$\times$10\textsuperscript{4})  &-1.98$\pm$0.20\textsuperscript{$\dagger$}&7.3e4 [7.5e4] (2.7e4)
(STB: +4$^{\circ}$)&b, d, e, g  \\
*19 &2012/03/07 &00:20&N17E027 [X5.4]&2700&71&3303& 13 &0.31&2.2$\times$10\textsuperscript{7} (1.9$\times$10\textsuperscript{4})  &-2.02$\pm$0.26\textsuperscript{$\dagger$}& 2.1e4 [6.0e4] (3.7e3) (STB: +14$^{\circ}$)&b, c, d, e, g   \\
*20&2012/05/17&01:30&N11W076 [M5.1]&1458&75&1807&-76&0.35&8.7$\times$10\textsuperscript{4} (2.0$\times$10\textsuperscript{4})&-& 3.6e2 [5.8e2] (1.1e1) (STA: -22$^{\circ}$)&b, d, e, f, g\\
*21&2012/05/26&20:40&N15W116 [-]&1850&55&2665&-75&0.31&1.9$\times$10\textsuperscript{4} (4.0$\times$10\textsuperscript{3})&-1.70$\pm$0.53& 1.3$\times$10\textsuperscript{4} [7.2$\times$10\textsuperscript{3}] (2.3e1) (STA: +22$^{\circ}$)&b, c, f, g\\
*22&2012/05/27&05:10\textsuperscript{ˆ}&S10E054 [C3.1]&1052&78&958&108&0.31&1.3$\times$10\textsuperscript{5} (2.4$\times$10\textsuperscript{4})&-2.56$\pm$0.96\textsuperscript{$\dagger$}& 1.2e5 [1.7e5] (8.7e3) (STA: +23$^{\circ}$)&-\\
*23&2012/07/12&15:45\textsuperscript{ˆ}&S15W001 [X1.4]&1393&75&1617& 4 &0.46&1.1$\times$10\textsuperscript{6} (5.5$\times$10\textsuperscript{3})  &-1.95$\pm$0.27\textsuperscript{$\dagger$}&-&b, d, f  \\
24&2012/07/17&14:00\textsuperscript{ˆ}&S20W065 [C9.9]&821&50&1245&59&0.46&1.6$\times$10\textsuperscript{4} (2.8$\times$10\textsuperscript{3})&-&-&b, f, g \\ 
25&2012/07/19&05:20&S13W088 [M7.7]&1500&71&1897&79&0.46&2.6$\times$10\textsuperscript{4} (7.1$\times$10\textsuperscript{3})&-&-&b, g\\
*26&2012/07/23&02:10\textsuperscript{ˆ}&S17W132 [-]&1900&116&2520&116&0.45&5.8$\times$10\textsuperscript{4} (9.5$\times$10\textsuperscript{3})&-1.90$\pm$0.18\textsuperscript{$\dagger$}&-&b, d, e, g \\
27&2012/07/28&21:05&S25E055 [M6.1]&792&68&1255&-82&0.44&5.4$\times$10\textsuperscript{4} (4.7$\times$10\textsuperscript{3})&-2.11$\pm$0.42\textsuperscript{$\dagger$}&-&-\\
*28 &2012/09/20 &14:55&S15E155 [-]&2600&54&3353& -29 &0.42&2.0$\times$10\textsuperscript{6} (2.5$\times$10\textsuperscript{4})  &-1.91$\pm$0.21\textsuperscript{$\dagger$}&-&b, d, e, g  \\
*29&2012/10/14&00:35&N13E137 [-]&1200&61&1502&-58&0.46&1.9$\times$10\textsuperscript{5} (4.0$\times$10\textsuperscript{3})&-1.93$\pm$0.15\textsuperscript{$\dagger$}
& 6.9e1 [1.3e2] (4.0e1)
(STB: -25$^{\circ}$)&b, g\\
30 &2013/03/16 &05:45&
S15W045 [C2.8]&260&61&-& -14 &0.43&2.7$\times$10\textsuperscript{5} (5.0$\times$10\textsuperscript{4}) &-1.92$\pm$0.45\textsuperscript{$\dagger$}&Previous event bg. (ACE:-3$^{\circ}$)&-  \\
*31&2013/04/11&07:00&N09E012 [M6.5]&1350&130&1602&-122&0.46&2.2$\times$10\textsuperscript{4} (2.7$\times$10\textsuperscript{3})&-&-&b, d, f \\
32 &2013/04/24 &21:40&N10W175 [-]&560&73&1017& 38 &0.40&3.3$\times$10\textsuperscript{6} (7.6$\times$10\textsuperscript{3})  &-2.22$\pm$0.16\textsuperscript{$\dagger$}&-&b, d \\
*33 &2013/05/13 &15:55&N11E085 [X2.8]&2000&84&2308& 67 &0.31&2.4$\times$10\textsuperscript{4} (6.3$\times$10\textsuperscript{3})&-1.80$\pm$0.59& 4.1e2 [6.0e2] (9.2e1) (STA: +13$^{\circ}$)&b, g\\
*34 &2013/06/21 &02:50\textsuperscript{ˆ}&S16E073 [M2.9]&1428&60&2303& -67 &0.46&5.5$\times$10\textsuperscript{5} (4.7$\times$10\textsuperscript{3})  &-1.82$\pm$0.30\textsuperscript{$\dagger$}&-& b, f, g \\
35 &2013/08/19 &01:20\textsuperscript{ˆ}&N10W162 [-]&-&-&-& -13 &0.32&4.0$\times$10\textsuperscript{4} (1.5$\times$10\textsuperscript{4})&-& No maximum (STA: -29$^{\circ}$)&-\\
*36 &2013/08/19 &22:30&N08W178 [M3.3\textsuperscript{§}]&1149&118&1192&-1 &0.32&2.9$\times$10\textsuperscript{7} (1.0$\times$10\textsuperscript{4})  &-1.99$\pm$0.25\textsuperscript{$\dagger$}
& 4.1e4 [8.6e4] (9.6e2) (STA: -24$^{\circ}$)&b, d, f \\
*37 &2013/10/11 &07:10&N21E103 [M1.5]&875&160&1267&-56 &0.43&1.4$\times$10\textsuperscript{5} (4.6$\times$10\textsuperscript{3})  &-1.92$\pm$0.08\textsuperscript{$\dagger$}&5.9e3 [2.1e3] (4.4e1)
 (STB: +28$^{\circ}$)&b, d, e, g \\
*38&2013/10/25&08:00&S10E073 [X1.7]&500&65&1188&-62&0.36&2.2$\times$10\textsuperscript{5} (1.3$\times$10\textsuperscript{4})&-1.85$\pm$0.16\textsuperscript{$\dagger$}&-&b,d, e \\
*39&2013/10/25&15:00&S06E069 [X2.1]&1225&69&1686&-59&0.36&2.8$\times$10\textsuperscript{5} (5.4$\times$10\textsuperscript{4})&-1.89$\pm$0.18\textsuperscript{$\dagger$}&-&b, g, f\\
*40 &2013/10/28 &15:10&S08E028 [M4.4]&1400&56&1393&-29 &0.34&8.1$\times$10\textsuperscript{5} (2.1$\times$10\textsuperscript{4})  &-1.97$\pm$0.06\textsuperscript{$\dagger$}&-&b, d, e \\
*41&2013/11/19&10:25&S15W069 [X1.0]&1138&52&1361&-41&0.34&6.2$\times$10\textsuperscript{4} (5.4$\times$10\textsuperscript{4})&-1.93$\pm$0.31\textsuperscript{$\dagger$}&-&b\\
*42&2013/11/30&05:10\textsuperscript{ˆ}&N13W150[-]&-&-&-&2&0.40&1.5$\times$10\textsuperscript{4} (4.9$\times$10\textsuperscript{3})&-
&-&b\\
*43&2013/11/30&15:00\textsuperscript{ˆ}&S15E146 [-]&830&48&830&65&0.40&1.6$\times$10\textsuperscript{4} (8.2$\times$10\textsuperscript{3})&-&-&b\\
*44&2013/12/26&03:05&S09E166 [-]&1738&47&1753&-9&0.46&1.1$\times$10\textsuperscript{6} (4.2$\times$10\textsuperscript{3})&-2.02$\pm$0.38\textsuperscript{$\dagger$}& 1.6e4 [2.0e4] (2.4e1) (STA: -6$^{\circ}$)&b, d, f, g \\
*45&2014/01/07&18:05&S15W011 [X1.2]&2190&61&2486&145&0.43&3.2$\times$10\textsuperscript{4} (6.1$\times$10\textsuperscript{3})&-& Several events mixed (STA: +16$^{\circ}$)&d, e, f, g\\
*46 &2014/01/28 &00:30\textsuperscript{ˆ}&S10E081 [C7.6] &-&-&-&-8 & 0.32 &5.9$\times$10\textsuperscript{3} (8.1$\times$10\textsuperscript{2})&-& Ion contamination
 (STB: +17$^{\circ}$)&- \\
47 &2014/01/28 &05:25\textsuperscript{ˆ}&S14E088 [C9.3] &-&-&-&-16 & 0.32 &2.2$\times$10\textsuperscript{4} (2.7$\times$10\textsuperscript{3})&-2.02$\pm$1.02\textsuperscript{$\dagger$}& Ion contamination (STB:+18$^{\circ}$)&- \\
48 &2014/01/30 &16:05 &S13E058 [M6.6] &1450&66&1367&2 & 0.31 &7.4$\times$10\textsuperscript{4} (7.1$\times$10\textsuperscript{3})&-1.82$\pm$0.33\textsuperscript{$\dagger$}&-&g\\
49 &2014/02/20 &07:50&S15W073 [M3.0] &1103&70&1328&34 & 0.37 &1.3$\times$10\textsuperscript{4} (1.5$\times$10\textsuperscript{3})&-& 1.3e4 [2.6e3] (6.7e2)\textsuperscript{\&} (ACE: -22$^{\circ}$)&g \\
*50 &2014/02/25 &00:45&S12E082 [X4.9]&2350&69&2431 &-137 &0.40 &5.5$\times$10\textsuperscript{4} (1.2$\times$10\textsuperscript{3})&-1.91$\pm$0.47\textsuperscript{$\dagger$}& 6.7e3 [7.2e3] (1.2e2)\textsuperscript{\&}
 (ACE: -6$^{\circ}$)&d, e, f, g\\
*51 &2014/03/13 &21:40\textsuperscript{ˆ}&N15W140 [-] &498&23&803&44 &0.46 &2.3$\times$10\textsuperscript{4} (3.8$\times$10\textsuperscript{3})&-1.55$\pm$0.31&2.6e2 [5.3e2] (1.4e2)\textsuperscript{\&}(ACE: +35$^{\circ}$)&-\\
52&2014/08/08&16:15&S10W160 [-]&1035&57&1352&-41&0.33&7.3$\times$10\textsuperscript{4} (6.2$\times$10\textsuperscript{3})&-1.82$\pm$0.21\textsuperscript{$\dagger$}& 1.0e2 [2.1e2] (4.6e1) (STA: -23$^{\circ}$)&g\\
*53 &2014/09/01 &11:00&N14E127 [-]&1842&77&2947&-44 &0.45&2.9$\times$10\textsuperscript{7} (3.4$\times$10\textsuperscript{3})  &-1.81$\pm$0.03\textsuperscript{$\dagger$}& 4.5e5 [2.5e5] (1.4e2)
 (STB: +15$^{\circ}$)&d, e, g \\
54 &2014/09/05 &06:50&S14E069 [C6.8]&565\textsuperscript{!}&56\textsuperscript{!}&NP& 6 &0.46&8.6$\times$10\textsuperscript{4} (3.9$\times$10\textsuperscript{4})&-2.06$\pm$0.65&No SEE (STB: +23$^{\circ}$)&- \\
55 &2014/09/08 &23:55&N12E029 [M4.5]&1120&36&1077& 39 &0.47&2.6$\times$10\textsuperscript{4} (5.4$\times$10\textsuperscript{3})&-& No SEE (STB: +30$^{\circ}$)&g \\
*56 &2014/09/10 &17:30&N14E002 [X1.6]&1580&74&1427& 64 &0.47&5.6$\times$10\textsuperscript{4} (1.0$\times$10\textsuperscript{4})&-1.77$\pm$0.16\textsuperscript{$\dagger$}& 9.3e2 [2.2e3] (3.1e2)
 (STB: +32$^{\circ}$)&d, g \\
*57 &2014/09/24 &20:45&N13E179 [-]&1516&76&1651&-139 &0.44&5.3$\times$10\textsuperscript{4} (4.7$\times$10\textsuperscript{3})  &-2.19$\pm$0.13\textsuperscript{$\dagger$}&-&d, g \\
58 &2014/12/13 &14:05\textsuperscript{ˆ}&S20W143 [-]&2036\textsuperscript{!}&92\textsuperscript{!}&2519\textsuperscript{!}&-75 &0.46&7.8$\times$10\textsuperscript{6} (3.4$\times$10\textsuperscript{3})  &-1.92$\pm$0.26\textsuperscript{$\dagger$}& No data (STA: -13$^{\circ}$)&d, g \\
59&2015/02/21&09:30\textsuperscript{ˆ}&S40W075 [B4.8]&884\textsuperscript{!}&65\textsuperscript{!}&1088\textsuperscript{!}&-19&0.44&3.8$\times$10\textsuperscript{4} (3.9$\times$10\textsuperscript{3})&-&1.2e3 [2.7e3] (1.8e2)\textsuperscript{\&} (ACE:+33$^{\circ}$)&-\\
60 &2015/03/24 &08:30\textsuperscript{ˆ}&S01W121 [-]&1371\textsuperscript{!}&106\textsuperscript{!}&2102\textsuperscript{!}&-31 &0.43&1.2$\times$10\textsuperscript{6} (1.3$\times$10\textsuperscript{4})  &-1.94$\pm$0.24\textsuperscript{$\dagger$}&-&- \\
*61 &2015/04/14 &09:15\textsuperscript{ˆ}&S15W100 [B9] &484\textsuperscript{!}&31\textsuperscript{!}&NP &-119 & 0.32&1.5$\times$10\textsuperscript{4} (4.5$\times$10\textsuperscript{3})&-&-&-\\

 \hline

\end{longtable}

\end{ThreePartTable}

\twocolumn 

\end{appendix}

\end{document}